\documentclass[sigconf]{acmart}

\usepackage{tabularx}
\usepackage{booktabs}
\usepackage{makecell}
\usepackage{array}
\usepackage{adjustbox}
\usepackage{graphicx}
\usepackage{xcolor}
\usepackage{tcolorbox}
\tcbuselibrary{skins,breakable}
\usepackage{subcaption}
\usepackage{float}

\usepackage{amssymb}
\usepackage{colortbl}
\usepackage{pifont}
\newcommand{\cmark}{\ding{51}}
\newcommand{\xmark}{\ding{55}}
\newcommand{\xhdr}[1]{\vspace{2mm}\noindent{{\bf #1.}}}
\usepackage{amsthm}
\usepackage{enumitem}
\usepackage{wrapfig}
\usepackage{multirow}

\usepackage{listings}
\AtBeginDocument{%
  }

\setcopyright{none}
\renewcommand\footnotetextcopyrightpermission[1]{}
\begin{document}

\title{SWE-Bench Mobile: Can Large Language Model Agents Develop Industry-Level Mobile Applications?}

\author{%
  Muxin Tian\textsuperscript{1,2,*} \quad
  Zhe Wang\textsuperscript{2,*} \quad
  Blair Yang\textsuperscript{3} \quad
  Zhenwei Tang\textsuperscript{1} \quad
  Kunlun Zhu\textsuperscript{4} \\
  Honghua Dong\textsuperscript{1} \quad
  Hanchen Li\textsuperscript{5} \quad
  Xinni Xie\textsuperscript{2} \quad
  Guangjing Wang\textsuperscript{2} \quad
  Jiaxuan You\textsuperscript{4,\textdagger}%
}
\affiliation{%
  \institution{%
    \textsuperscript{1}University of Toronto \quad
    \textsuperscript{2}Xiaohongshu Inc. \quad
    \textsuperscript{3}Coolwei AI Lab \quad
    \textsuperscript{4}University of Illinois Urbana-Champaign \\
    \textsuperscript{5}University of California, Berkeley%
  }%
  \country{}%
}
\renewcommand{\shortauthors}{Tian et al.}

\begin{abstract}
Can large language model agents develop industry-level mobile applications? We introduce \textbf{SWE-Bench Mobile}, a benchmark for evaluating coding agents on realistic software engineering tasks derived from a production iOS codebase. Unlike existing benchmarks that focus on isolated problems or bug fixes, SWE-Bench Mobile captures the full complexity of industrial development: multi-modal inputs (PRDs and Figma designs), a large-scale mixed Swift/Objective-C codebase, and comprehensive test suites. We evaluate 22 agent-model configurations across four coding agents---three commercial (Cursor, Codex, Claude Code) and one open-source (OpenCode)---and find that even the best configurations achieve only 12\% task success rate. Our analysis reveals that (1) agent design matters as much as model capability---the same model shows up to 6$\times$ performance gap across agents, (2) commercial agents consistently outperform open-source alternatives, and (3) simple ``Defensive Programming'' prompts outperform complex ones by 7.4\%. These findings highlight a significant gap between current agent capabilities and industrial requirements, while providing actionable insights for practitioners and researchers. We release SWE-Bench Mobile as a \textit{hosted benchmark challenge} to prevent data contamination and ensure fair evaluation. The public leaderboard and development toolkit are available at \url{https://swebenchmobile.com}.
\end{abstract}


\maketitle
\pagestyle{plain}
\fancyhead{}

\begingroup
\renewcommand{\thefootnote}{}
\footnotetext{\textsuperscript{*}Equal contribution. \textsuperscript{\textdagger}Corresponding author: \texttt{jiaxuan@illinois.edu}.}
\endgroup

\section{Introduction}

\begin{figure*}[t]
    \centering
    \includegraphics[width=\textwidth]{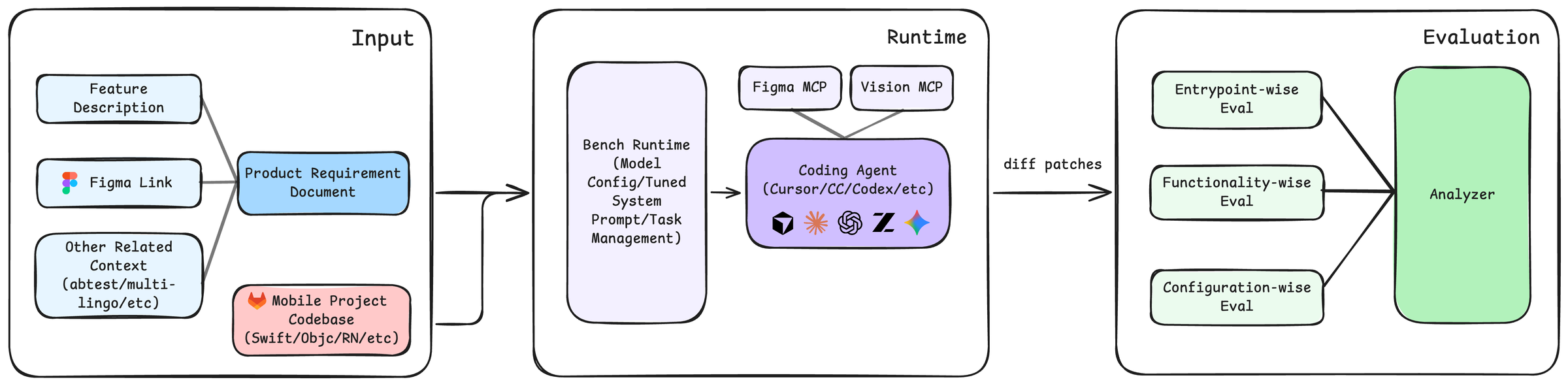}
    \caption{Overview of the SWE-Bench Mobile pipeline. (1) Agents receive multi-modal inputs including a Product Requirement Document (PRD), Figma design, and a large-scale Swift/Objective-C codebase. (2) The agent navigates the codebase, plans the implementation, and generates code. (3) The output is a Git patch that is applied and evaluated against a comprehensive test suite.}
    \label{fig:pipeline}
\end{figure*}

Large language models~(LLMs) have enabled a new generation of autonomous coding agents that can understand requirements, navigate codebases, and implement features with minimal human intervention. Commercial systems like GitHub Copilot, Cursor, and Claude Code have achieved impressive results on existing benchmarks, raising a critical question: \textit{Can these agents handle the complexity of real-world, industry-level mobile software development?}

Answering this question requires a comprehensive evaluation that faithfully captures professional software engineering. However, existing benchmarks have significant limitations. HumanEval~\cite{humaneval} and MBPP~\cite{mbpp} evaluate isolated algorithmic problems far removed from industrial practice. SWE-Bench~\cite{swebench} advances the field by using real GitHub issues, but still falls short of industrial realism: it focuses on bug fixes rather than feature development, uses text-only inputs without design specifications, typically involves small localized changes to 1-2 files, and concentrates on Python which is well-represented in training data. Recent work like SWE-Bench Pro~\cite{swebenchpro} addresses some limitations by introducing longer-horizon tasks, but still lacks multi-modal inputs and focuses exclusively on Python.

In professional software development, engineers participate in a structured workflow that goes far beyond writing code. They interpret Product Requirement Documents (PRDs) that specify what to build and why. They translate visual designs from tools like Figma into implementation decisions about layout and interaction. They navigate large codebases---often hundreds of thousands of lines---to find relevant files and understand existing patterns. They make coordinated changes across multiple modules while maintaining consistency. And they ensure their implementations handle edge cases and pass comprehensive tests. A benchmark claiming to evaluate ``industry-level'' capabilities must test all of these aspects.

We focus on mobile application development not merely for language diversity, but because it represents a distinct and critical paradigm in software engineering that remains unexplored by current benchmarks. Unlike server-side logic (e.g., Python scripts), mobile development introduces unique challenges for AI agents:
(1) \textbf{Multi-modal Dependency:} Implementation is strictly guided by visual designs (Figma) and user interactions, requiring agents to perform visually-grounded program synthesis rather than just text-to-code generation.
(2) \textbf{Event-Driven Complexity:} Mobile apps are stateful systems that must handle asynchronous user events, network changes, and strict OS lifecycle callbacks, challenging agents' ability to model dynamic system states.
(3) \textbf{Client-Side Constraints:} Development occurs within framework-heavy environments (e.g., iOS SDK) with rapid iterations, testing generalization to domain-specific APIs.

We introduce \textbf{SWE-Bench Mobile}, a benchmark for evaluating coding agents on industry-level mobile application development. SWE-Bench Mobile is constructed from real development artifacts at a major technology company, comprising 50 authentic tasks derived from actual product requirements. Each task combines multi-modal inputs---PRDs, Figma designs, and a large-scale mixed Swift/Objective-C production codebase---with comprehensive evaluation through 449 human-verified test cases.

\xhdr{Contributions}
\begin{enumerate}[leftmargin=*]
    \item We introduce SWE-Bench Mobile, the first benchmark combining PRDs, Figma designs, and a large-scale codebase to capture the full complexity of industrial software development.
    \item We evaluate 22 agent-model configurations across four coding agents (three commercial, one open-source), with detailed analysis of performance, cost, and robustness.
    \item We systematically categorize agent failures, finding that 54\% stem from missing feature flags---a production practice unfamiliar to agents---followed by missing data models (22\%) and incomplete file coverage (11--15\%).
    \item We provide actionable insights: agent design matters as much as model capability (up to 6$\times$ performance gap for the same model), commercial agents outperform open-source ones, simple prompts outperform complex ones, and cost-effective configurations exist.
\end{enumerate}

To strictly preserve the integrity of the evaluation and respect the proprietary nature of the production codebase, we adopt a \textit{hosted evaluation paradigm}. Unlike static datasets that are prone to data contamination in future model training sets, our \textit{held-out private test set} ensures that agents are evaluated on truly unseen industrial tasks. We provide a \textit{sanitized development kit} and a public leaderboard to foster community progress.

Our evaluation reveals a significant gap between current capabilities and industrial requirements. The best configuration achieves only 12\% task success rate, with most failures due to incomplete implementations. The same model (Opus 4.5) achieves 12\% on Cursor but only 2\% on OpenCode---a 6$\times$ gap---demonstrating that agent scaffolding matters as much as model capability. Commercial agents consistently outperform the open-source OpenCode, whose best result (8\% with GLM 4.6) trails the best commercial result (12\%) by 4 percentage points. Success drops from 18\% for simple tasks requiring 1-2 files to just 2\% for complex tasks requiring 7+ files, indicating agents struggle with cross-file reasoning. These findings suggest that while coding agents show promise for simple tasks, substantial improvements in requirement understanding, multi-modal reasoning, and codebase navigation are needed for reliable industry-level development.

\section{SWE-Bench Mobile}

SWE-Bench Mobile is a benchmark designed to evaluate coding agents on industry-level mobile application development. Unlike existing benchmarks that focus on isolated coding problems or bug fixes, SWE-Bench Mobile captures the full complexity of professional software engineering: multi-modal inputs, large codebases, and comprehensive testing. Figure~\ref{fig:pipeline} illustrates the overall benchmark pipeline.

\subsection{Problem Formulation}

Each benchmark instance is represented as a triplet:
\[
\mathcal{T} = (\mathcal{I}, \mathcal{O}, \mathcal{E}),
\]
where $\mathcal{I}$ is the input context, $\mathcal{O}$ is the expected output, and $\mathcal{E}$ is the evaluation configuration.

\begin{figure*}[t]
    \centering
    \includegraphics[width=\textwidth]{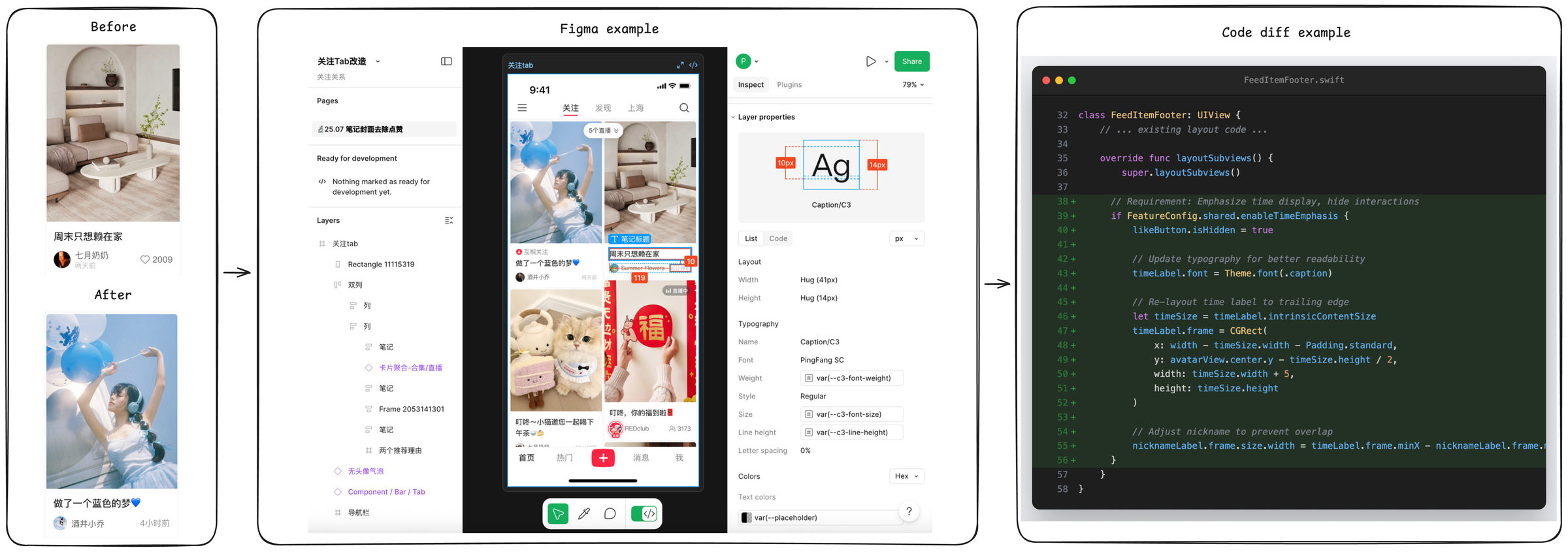}
    \caption{A concrete example of a SWE-Bench Mobile task (Task 056). The agent must interpret the PRD requirements (replace interaction button with publish time label) and visual design (Figma), locate the relevant files in the codebase (FeedItemFooter.swift), and implement the changes while handling edge cases and feature configuration.}
    \label{fig:task_example}
\end{figure*}

\xhdr{Input ($\mathcal{I}$)}
The input context mimics a typical developer's starting point for a new feature. It consists of three main components (see Figure~\ref{fig:task_example}). First, a \textbf{Product Requirement Document (PRD)} describes the feature goal, user story, acceptance criteria, and constraints. These PRDs are derived from actual product requirements at XiaoHongShu Inc., a major social media platform with over 300 million monthly active users, follow standard industrial conventions~\cite{atlassian_prd}, and have an average length of 450 words, requiring agents to parse natural language specifications. Second, 70\% of tasks include a \textbf{Figma Design} specification, containing component layout, typography, and visual details that the agent must translate into code. Finally, the agent is provided with the \textbf{XiaoHongShu production codebase}, a Git repository snapshot containing approximately 500,000 lines of Swift/Objective-C code across thousands of files. This large-scale context forces the agent to perform retrieval and navigation, rather than just code generation.

\xhdr{Output ($\mathcal{O}$)}
The expected output is a unified diff patch that, when applied to the codebase, implements the feature described in the PRD. This format matches the standard pull request workflow used in industry.

\xhdr{Evaluation ($\mathcal{E}$)}
Each task is paired with a task-specific \texttt{pytest} suite (9.1 tests per task on average) that evaluates the generated \emph{patch} directly. Concretely, tests operate on the unified diff \emph{text} without compiling or running the iOS application, and therefore avoid build-time overhead and simulator/device nondeterminism. This patch-level evaluation is designed to verify the \emph{presence} of necessary UI-facing edits (e.g., view construction, layout logic) and data/logic edits (e.g., control-flow, state updates), while remaining tolerant to superficial variability such as identifier naming, refactoring style, and minor structural reorganization.

\subsection{Design Principles}

\begin{figure}[t]
\centering
\includegraphics[width=\columnwidth]{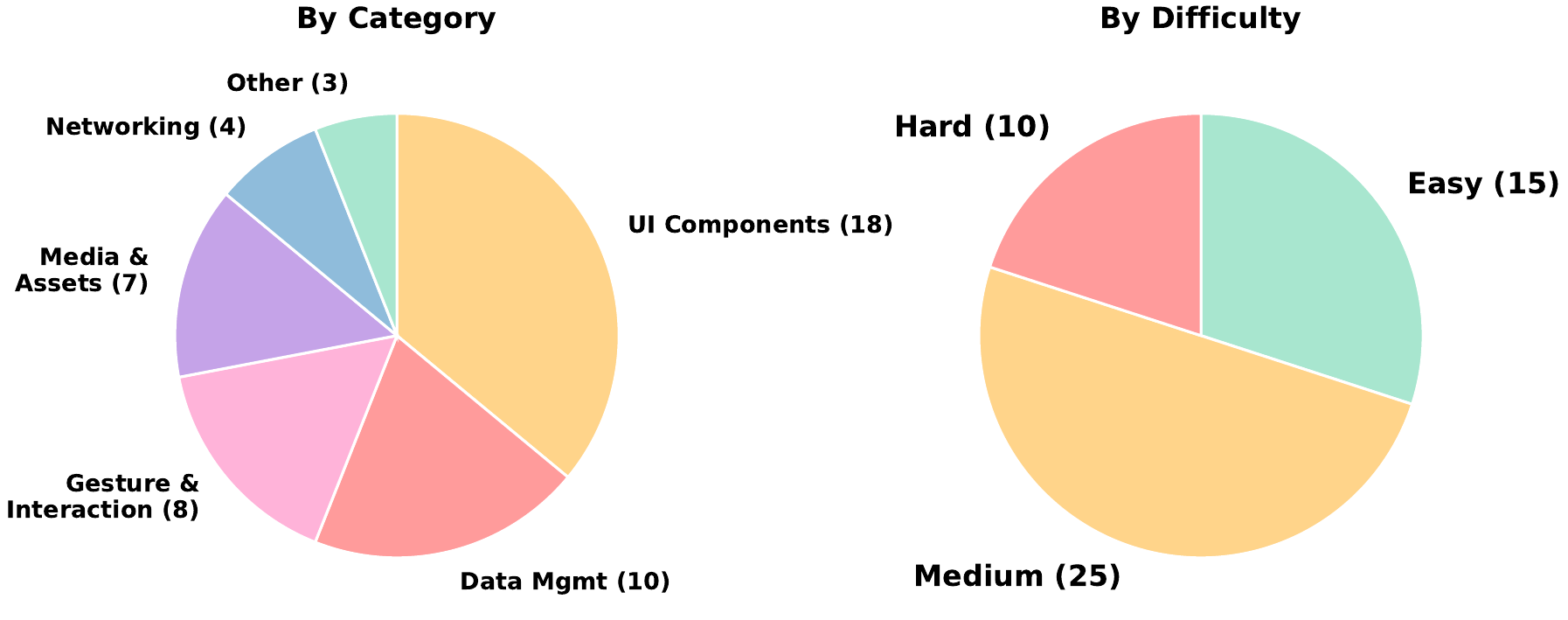}
\caption{Task distribution by category (left) and difficulty (right). Each label shows the count, percentage, and average agent pass rate. UI Components (36\%) dominate the benchmark, while performance drops sharply from Easy (18.5\% pass) to Hard (5.8\% pass).}
\label{tab:task_categories}
\end{figure}

SWE-Bench Mobile is constructed under guiding principles to ensure relevance to professional software engineering. \textbf{End-to-End Realism} is paramount; tasks span the full engineering process from PRD to testing, preserving real-world dependencies and incomplete specifications. Unlike synthetic benchmarks, our tasks come from actual product development cycles. \textbf{Multi-Modal Reasoning} is required, as agents must jointly interpret textual requirements (PRD), visual designs (Figma), and structured code. \textbf{Diverse Coverage} ensures robustness, with tasks covering multiple categories (Table~\ref{tab:task_categories}) and difficulty levels, from simple UI adjustments to complex architectural refactoring. Finally, by focusing on Swift/Objective-C, an \textbf{Under-Represented Language} in LLM training data compared to Python or JavaScript, SWE-Bench Mobile serves as a challenging test of an agent's ability to generalize to less familiar syntax and frameworks.

\subsection{Dataset Statistics}

Table~\ref{tab:dataset_stats_main} summarizes the key statistics of SWE-Bench Mobile. The benchmark consists of 50 tasks with 449 total test cases. The majority of tasks (70\%) include Figma designs, and 92\% include reference images, highlighting the multi-modal nature of the dataset. The average PRD length is 450 words, providing substantial context. The codebase scale is significant, with the repository size reaching approximately 5GB.

\begin{table}[t]
\centering
\caption{SWE-Bench Mobile dataset statistics.}
\label{tab:dataset_stats_main}
\small
\begin{tabular}{lr}
\toprule
\textbf{Metric} & \textbf{Value} \\
\midrule
Total Tasks & 50 \\
Total Test Cases & 449 \\
Avg. Test Cases per Task & 9.1 \\
\midrule
Tasks with Figma Design & 35 (70\%) \\
Tasks with Reference Images & 46 (92\%) \\
Avg. PRD Length (words) & 450 \\
\midrule
Codebase Size & Large Scale ($\sim$5GB) \\
Programming Language & Swift/Objective-C (iOS) \\
Avg. Files Modified per Task & 4.2 \\
\bottomrule
\end{tabular}
\end{table}

\subsection{Task Construction}

\xhdr{Source}
Tasks are derived from real product requirements at XiaoHongShu Inc., a leading social media platform in China with over 300 million monthly active users. Each task represents a feature that was actually implemented by XiaoHongShu engineers in the production iOS application, ensuring realistic complexity and scope. Unlike existing benchmarks that use synthetic problems or isolated bug fixes from open-source repositories, our tasks capture the full complexity of feature development in a commercial mobile application: multi-file changes, UI/UX implementation from design specs, integration with existing business logic, and handling of edge cases and feature flags. This industry-sourced approach ensures that our benchmark reflects the actual challenges faced by software engineers in production environments.

\xhdr{Quality Control}
Each task undergoes a rigorous multi-stage review process. First, the PRDs are reviewed to ensure requirements are clear and self-contained. Next, comprehensive test suites are designed to verify both correctness and quality. Finally, we perform human validation to verify that the reference implementation passes all tests.

\xhdr{Difficulty Calibration}
Tasks are labeled by implementation complexity based on several factors: the number of files to modify (1-2 for Easy, 3-5 for Medium, 6+ for Hard), the lines of code changed ($<$50 for Easy, 50-150 for Medium, $>$150 for Hard), and the architectural complexity, distinguishing between localized changes and cross-module refactoring.

\subsection{Evaluation Pipeline}

Unlike traditional code benchmarks that rely solely on unit tests, SWE-Bench Mobile performs comprehensive verification through a multi-step pipeline.

\xhdr{Patch-to-Task Routing}
SWE-Bench Mobile evaluates submissions as unified diff patches and associates each patch with a specific task. This routing step ensures that each submission is evaluated under the task's PRD-defined intent and its corresponding test suite, while keeping the evaluation independent of repository checkout, compilation, or runtime execution. In practice, the test harness exposes the patch \emph{text} to the task-specific tests, enabling purely diff-based verification.

\xhdr{Static Analysis}
Before running task-specific assertions, we perform lightweight static checks on the diff text. This includes verifying unified diff structure (e.g., \texttt{diff --git} headers), rejecting empty or near-empty patches, and ensuring that added lines contain meaningful code changes rather than only whitespace or comments. We also check whether the patch touches relevant files using flexible path patterns (e.g., accepting file moves/renames), and apply basic language-agnostic sanity checks to filter malformed submissions early.

\xhdr{Diff-Based Intent Tests}
Direct runtime evaluation for mobile applications is challenging to scale. Unit tests are ill-suited for validating visual correctness, while end-to-end UI testing introduces substantial compilation overhead and environmental nondeterminism. To address these constraints, SWE-Bench Mobile adopts a \textbf{diff-based evaluation} strategy: our \texttt{pytest} suites inspect the patch diff and verify \emph{structural intent} and \emph{architectural compliance}. This allows us to evaluate high-level architectural decisions and requirement compliance at scale. Tests are constructed from the PRD and a human reference patch, emphasizing:
\begin{itemize}[leftmargin=1.5em, topsep=2pt, itemsep=2pt, parsep=0pt]
    \item \textbf{Goal-oriented checks}: verifying modification patterns (the ``what'') rather than exact code shape.
    \item \textbf{Feature entry points}: checking integration surfaces (e.g., routing, hooks).
    \item \textbf{Removal of blocking behavior}: ensuring constraints or legacy guards are lifted.
    \item \textbf{Cohesion across files}: verifying related edits across modules.
    \item \textbf{Semantics-aware matching}: using flexible pattern matching to accommodate alternative naming.
\end{itemize}

\xhdr{Batch Reporting and Error Analysis}
Beyond pass/fail decisions, our evaluator produces both task-level and test-case-level summaries. For large-scale runs, we classify failures into coarse categories (e.g., missing critical file edits, missing UI components, empty patches). This analysis provides interpretable diagnostics of common agent failure modes and supports systematic iteration on prompts and agent scaffolding.

\xhdr{Metrics}
We report two complementary metrics. \textbf{Task Success Rate} is the percentage of tasks where \textit{all} tests pass, representing the strict standard for a completed feature. \textbf{Test Pass Rate} is the percentage of individual test cases passed, which reveals partial progress even when the full task is not completed. The gap between these metrics reveals how often agents make partial progress without fully completing tasks.

\subsection{Comparison with Existing Benchmarks}

Table~\ref{tab:benchmark_comparison} compares SWE-Bench Mobile with existing coding benchmarks. SWE-Bench Mobile distinguishes itself by being multi-modal, including PRDs and Figma designs rather than just code or text descriptions. It operates on a large-scale codebase ($\sim$5GB), significantly larger than the individual repositories or snippets used in other benchmarks. Furthermore, it targets mixed Swift/Objective-C, which is under-represented in training data compared to Python, and focuses on feature implementation rather than bug fixing.

\begin{table}[t]
\centering
\caption{Comparison with existing benchmarks.}
\label{tab:benchmark_comparison}
\small
\setlength{\tabcolsep}{3pt}
\begin{tabular}{lccc}
\toprule
\textbf{Benchmark} & \textbf{Multi-Modal} & \textbf{Codebase} & \textbf{Language} \\
\midrule
HumanEval & \xmark & None & Python \\
MBPP & \xmark & None & Python \\
SWE-Bench & \xmark & Medium & Python \\
\textbf{SWE-Bench Mobile} & \cmark & Large & Swift/ObjC \\
\bottomrule
\end{tabular}
\end{table}

\section{Experiments}

We evaluate leading coding agents on SWE-Bench Mobile to answer several key research questions. First, we investigate how state-of-the-art coding agents perform on industry-level mobile development tasks (\textbf{RQ1}). Second, we analyze how task complexity affects agent performance (\textbf{RQ2}). Third, we examine the cost-performance trade-off (\textbf{RQ3}). Fourth, we assess the robustness of agent results across multiple runs (\textbf{RQ4}). Finally, we explore how prompt engineering affects performance (\textbf{RQ5}).

\subsection{Experimental Setup}

\xhdr{Agents and Models}
We evaluate four coding agents spanning commercial and open-source systems: \textbf{Cursor}, an AI-powered code editor with an agent mode; \textbf{Codex}, OpenAI's coding agent CLI; \textbf{Claude Code}, Anthropic's coding agent CLI; and \textbf{OpenCode}, an open-source coding agent. We test these agents with multiple backbone models including Claude Opus 4.5, Claude Sonnet 4.5, Claude Haiku, GLM 4.6, GLM 4.7, GPT 5, GPT 5.1, GPT 5.2, and Gemini 3 Pro, yielding 22 agent-model configurations in total.

\xhdr{Metrics}
We report two primary metrics: \textbf{Task Success Rate}, which is the percentage of tasks where \textit{all} test cases pass, and \textbf{Test Pass Rate}, which is the percentage of individual test cases passed. All rates are computed with a fixed denominator of 50 tasks and 449 test cases. When an agent fails to produce a patch for a task (e.g., due to timeout or error), the missing patch is counted as failing all associated tests.

\subsection{Main Results (RQ1)}

\begin{figure}[t]
    \centering
    \includegraphics[width=\columnwidth]{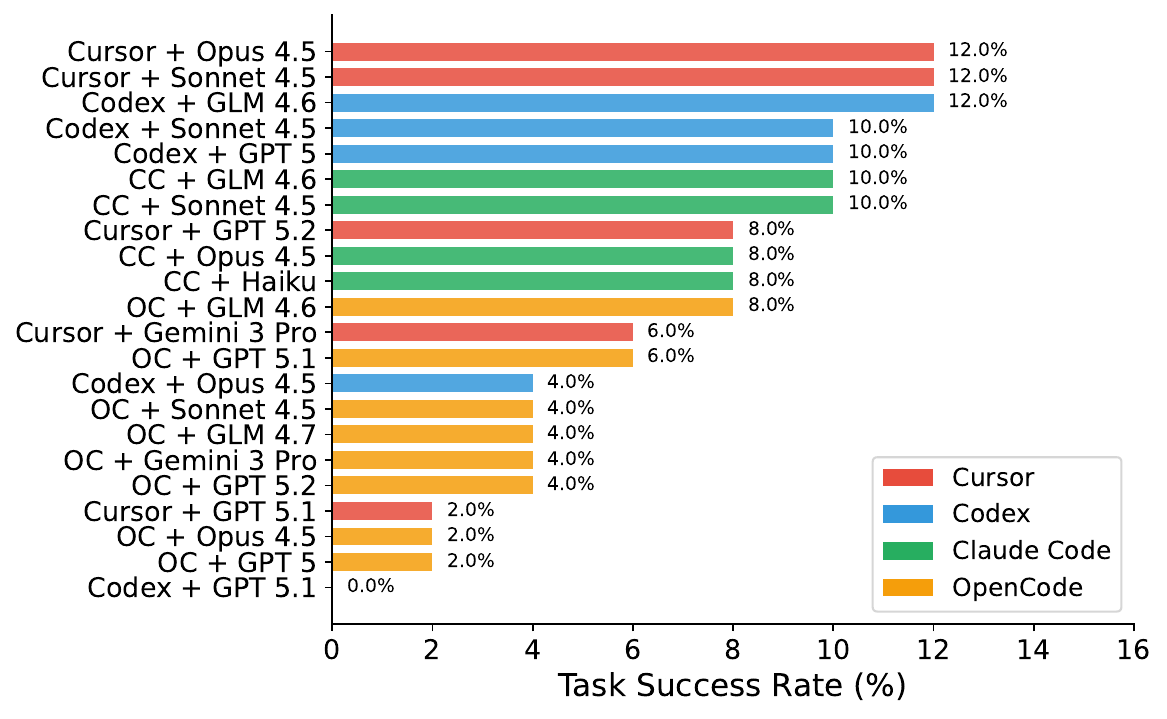}
    \caption{Task Success Rate across all configurations. Best performance is 12\%, achieved by Cursor + Opus/Sonnet and Codex + GLM.}
    \label{fig:main_results}
\end{figure}

Figure~\ref{fig:main_results} presents the main experimental results across all agent-model configurations.

\xhdr{Key Findings}
Our evaluation reveals a generally low overall performance, with even the best agents solving only 12\% of tasks. This indicates a significant gap between current capabilities and industrial requirements. However, the Test Pass Rate (up to 28.1\%) is much higher than the Task Success Rate (12\%), indicating that agents often make partial progress but fail to complete tasks fully. Notably, we find that the choice of agent matters significantly: the same model (Opus 4.5) achieves 12\% on Cursor but only 2\% on OpenCode, a 6$\times$ difference. Commercial agents consistently outperform the open-source OpenCode agent: the best OpenCode configuration (GLM 4.6, 8\%) trails the best commercial configuration (12\%) by 4 percentage points.

\subsection{Task Complexity Analysis (RQ2)}

\begin{figure}[t]
    \centering
    \includegraphics[width=\columnwidth]{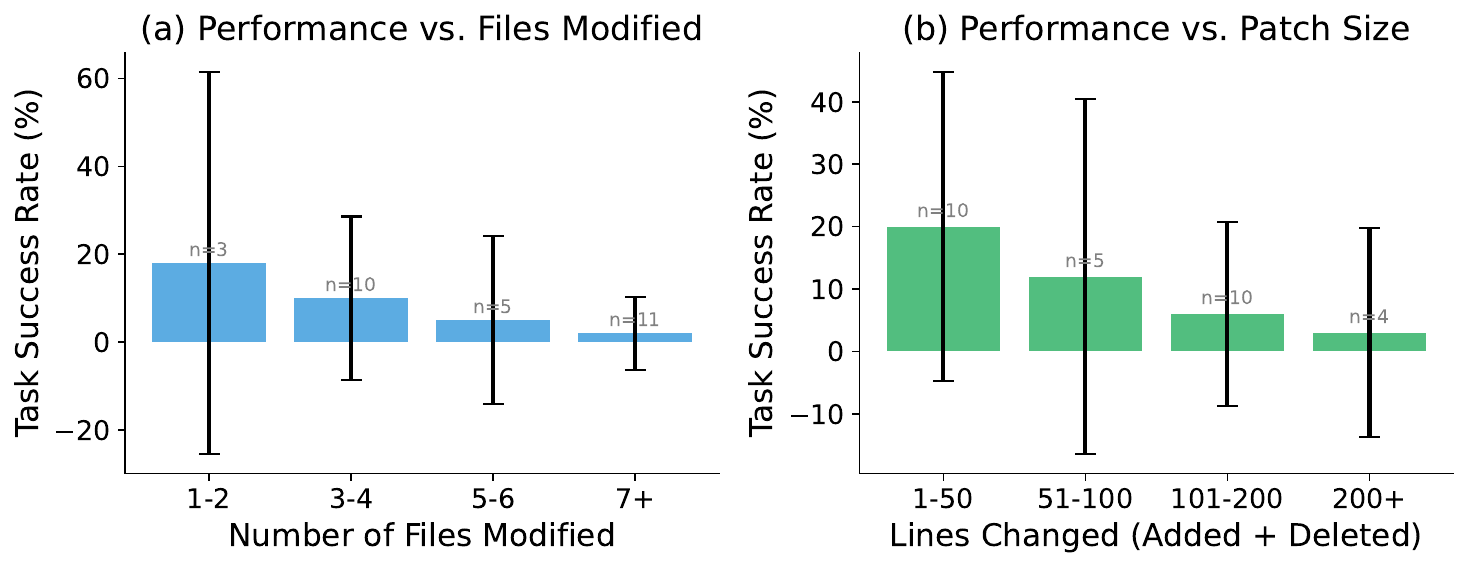}
    \caption{Performance decreases sharply with task complexity. (a) Tasks requiring 1-2 file modifications have 18\% success rate vs. 2\% for 7+ files. (b) Small patches ($<$50 lines) achieve 20\% success vs. 3\% for large patches ($>$200 lines). Error bars show 95\% confidence intervals based on binomial proportions.}
    \label{fig:complexity}
\end{figure}

We analyze how task complexity affects agent performance. Figure~\ref{fig:complexity} shows the relationship between task complexity (measured by number of files modified and patch size) and success rate.

\xhdr{Key Findings}
Performance drops sharply as complexity increases. The success rate drops from 18\% for tasks requiring 1-2 file modifications to just 2\% for tasks requiring 7+ files, suggesting that agents struggle with cross-file reasoning. Similarly, larger patches correlate with lower success, indicating difficulty with complex implementations.

\subsection{Model Comparison Across Agents}

\begin{figure}[t]
    \centering
    \includegraphics[width=\columnwidth]{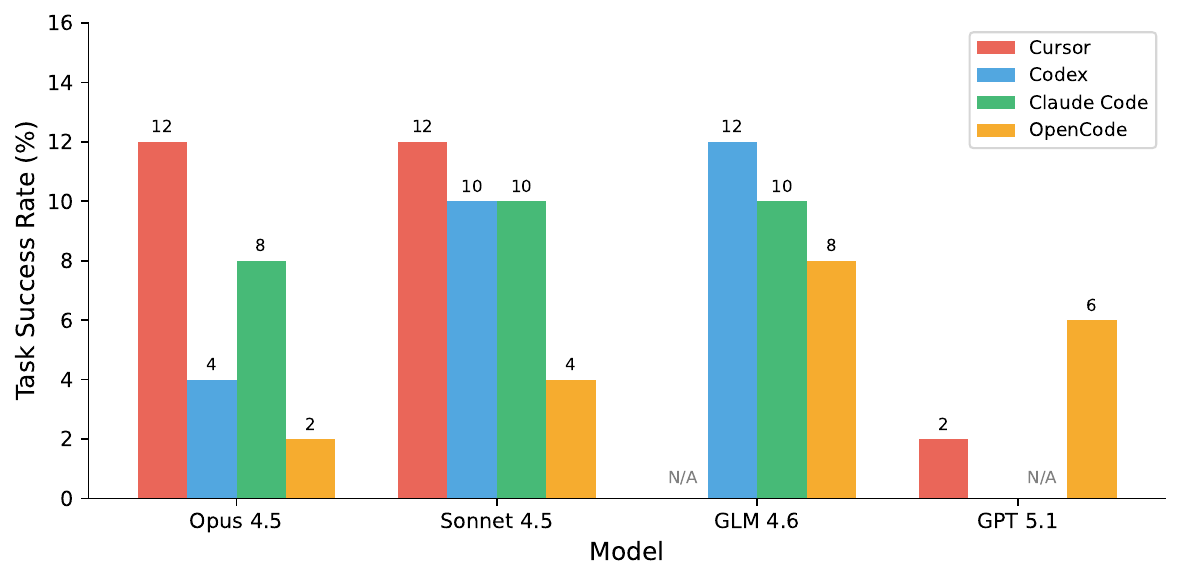}
    \caption{Same model, different agents: Opus 4.5 achieves 12\% on Cursor but only 2\% on OpenCode---a 6$\times$ gap. Commercial agents consistently outperform the open-source alternative.}
    \label{fig:model_comparison}
\end{figure}

A surprising finding is that \textbf{the same model performs very differently across agents}. Figure~\ref{fig:model_comparison} shows this comparison across all four agents.

\xhdr{Implications}
This finding suggests that agent scaffolding (tool use, context management, iteration strategy) is as important as the underlying model capability. The performance gap between commercial agents (Cursor, Codex, Claude Code) and the open-source OpenCode is substantial across all models, suggesting that years of engineering investment in tool integration, context management, and iterative refinement provide significant advantages. Practitioners should evaluate agents holistically rather than focusing solely on model benchmarks.

\subsection{Performance by Task Category}
\begin{figure}[t]
    \centering
    \includegraphics[width=\columnwidth]{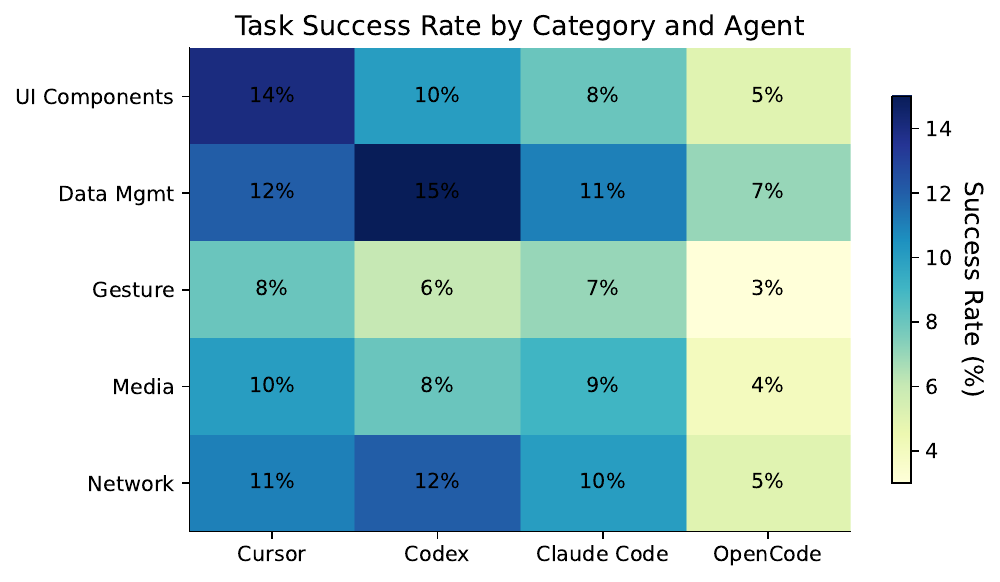}
    \caption{Task Success Rate by Category and Agent. Agents generally perform better on Data Management tasks but struggle with Gesture \& Interaction and Media tasks, which require complex multi-modal reasoning.}
    \label{fig:heatmap}
\end{figure}

We analyze how agents perform across different task categories. Figure~\ref{fig:heatmap} shows the success rate breakdown.

\subsection{Cost and Time Analysis (RQ3)}

Table~\ref{tab:cost_time} presents the cost and time metrics for each configuration. We measure API cost per task and average execution time.

\begin{table}[t]
\centering
\caption{Cost and time comparison across all agents. Best value in each column is \textbf{bold}. OpenCode costs are reported via OpenRouter API billing.}
\label{tab:cost_time}
\small
\setlength{\tabcolsep}{3pt}
\begin{tabular}{@{}llcc@{}}
\toprule
\textbf{Agent} & \textbf{Model} & \textbf{Cost (\$/task)} & \textbf{Time (min)} \\
\midrule
Cursor & Opus 4.5 & 3.50 & 15.0 \\
Cursor & Sonnet 4.5 & 2.00 & 14.2 \\
Codex & GLM 4.6 & 1.30 & 13.3 \\
Codex & Sonnet 4.5 & 2.50 & 12.5 \\
CC & GLM 4.6 & 1.30 & 11.7 \\
CC & Sonnet 4.5 & 2.00 & 13.3 \\
CC & Opus 4.5 & 4.00 & 15.0 \\
CC & Haiku & 0.50 & 8.3 \\
\midrule
OC & Opus 4.5 & 9.33 & 8.2 \\
OC & Sonnet 4.5 & 3.50 & 11.1 \\
OC & GLM 4.6 & 0.13 & 32.5 \\
OC & GLM 4.7 & 0.49 & 52.1 \\
OC & GPT 5 & 0.18 & 9.8 \\
OC & GPT 5.1 & \textbf{0.02} & \textbf{2.0} \\
OC & GPT 5.2 & 0.04 & 10.9 \\
OC & Gemini 3 Pro & 0.03 & 8.9 \\
\bottomrule
\end{tabular}
\end{table}
\xhdr{Key Findings}
Among commercial agents, Codex + GLM 4.6 offers the best value, achieving 12\% success at only \$1.30/task---the same success rate as Cursor + Opus 4.5 but at less than half the cost (\$3.50/task). OpenCode exhibits a striking cost--time trade-off: it is dramatically cheaper (GLM 4.6 at \$0.13/task vs.\ \$1.30 for Codex/CC), but GLM models run much slower (32--52 min vs.\ 11--13 min). OpenCode + Opus 4.5 is the most expensive configuration at \$9.33/task yet achieves only 2\% success, while OpenCode + GPT 5.1 is the cheapest at \$0.02/task but completes tasks in only 2 minutes on average---likely because it fails quickly on most tasks (6\% success, 7.1\% test pass rate).

\subsection{Robustness Analysis (RQ4)}

\begin{figure}[t]
    \centering
    \includegraphics[width=0.85\columnwidth]{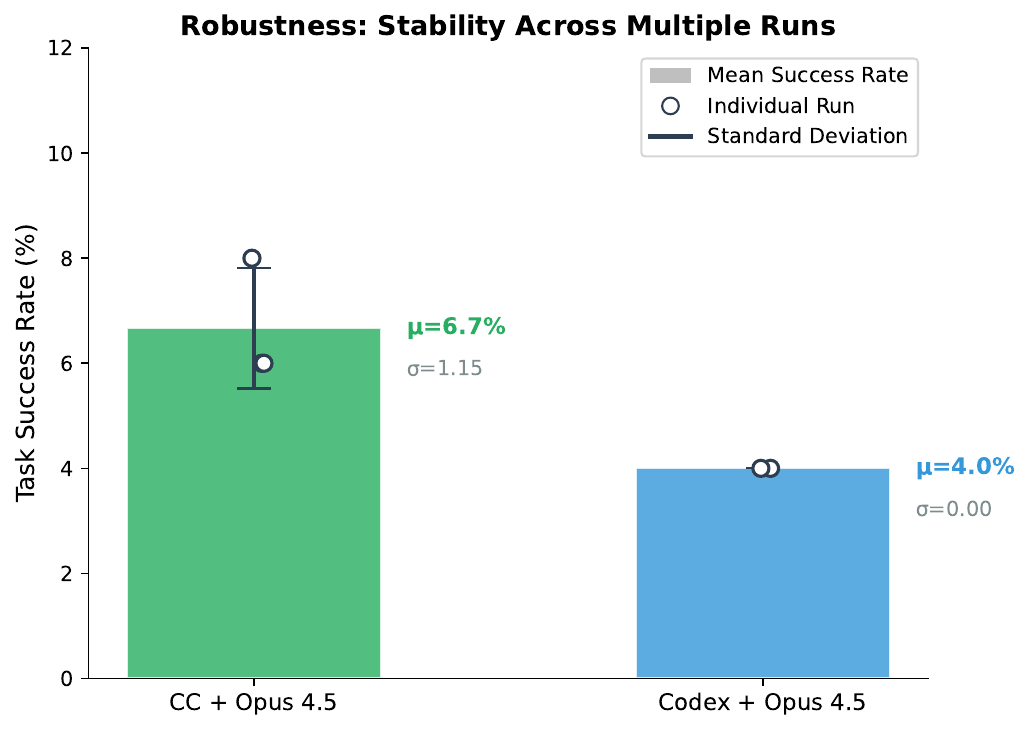}
    \caption{Result stability across multiple runs. Error bars indicate standard deviation. While Claude Code shows moderate variance ($\sigma$=1.15\%), the absolute fluctuation is small ($\pm$1 task), indicating that agent performance is relatively stable.}
    \label{fig:robustness}
\end{figure}

To assess result stability, we run selected configurations multiple times. Figure~\ref{fig:robustness} shows the variance across runs.

\xhdr{Observations}
We observe moderate variance for Claude Code + Opus 4.5, with scores of 6\%, 8\%, and 6\% across 3 runs ($\mu$=6.7\%, $\sigma$=1.15\%). In contrast, Codex + Opus 4.5 is perfectly stable at 4\% across runs.

\subsection{Prompt Engineering (RQ5)}
\label{sec:prompt_ablation}

We conduct a systematic ablation study with 12 prompt variants using Claude Code + GLM 4.6. Table~\ref{tab:prompt_ablation} shows the results.

\begin{table}[t]
\centering
\caption{Prompt ablation results. \colorbox{green!15}{Best} and \colorbox{red!15}{worst} highlighted. Full prompts in Appendix~\ref{app:prompt_ablation}.}
\label{tab:prompt_ablation}
\small
\setlength{\tabcolsep}{4pt}
\begin{tabular}{@{}lcc@{}}
\toprule
\textbf{Prompt Strategy} & \textbf{Task (\%)} & \textbf{Test (\%)} \\
\midrule
\rowcolor{green!15} Defensive Programming & \textbf{10.0} & \textbf{26.7} \\
Quality Focused & 10.0 & 26.3 \\
Example Driven & 10.0 & 23.4 \\
Chain of Thought & 10.0 & 21.8 \\
Baseline & 10.0 & 19.3 \\
\midrule
Explicit Instructions & 8.0 & 17.8 \\
Figma Emphasis & 8.0 & 18.0 \\
Test Driven & 6.0 & 22.0 \\
\midrule
\rowcolor{red!15} Detailed Role & 4.0 & 20.7 \\
\rowcolor{red!15} Structured Checklist & 4.0 & 20.7 \\
\rowcolor{red!15} Context Rich & 4.0 & 22.7 \\
\rowcolor{red!15} Comprehensive & 4.0 & 22.7 \\
\bottomrule
\end{tabular}
\end{table}

\xhdr{Key Findings}
The "Defensive Programming" prompt strategy performs best, improving the Test Pass Rate by 7.4\% over the baseline (19.3\% $\rightarrow$ 26.7\%) while maintaining the same Task Success Rate (10.0\%). This indicates that while both prompts complete the same number of tasks fully, Defensive Programming handles edge cases better in partially-completed tasks, passing significantly more individual test cases. This suggests that emphasizing defensive coding practices helps agents avoid common pitfalls even when they cannot complete all requirements. Interestingly, complexity appears to hurt performance; overly detailed prompts reduce Task Success from 10.0\% to 4.0\%. Overall, prompts focusing on code quality outperform those emphasizing workflow.

\subsection{Error Analysis}

We categorize failure modes across all experiments by analyzing test failure messages from the best-performing agents. The most critical failure pattern is \textbf{Missing Feature Flags} (54\%), where agents implement core functionality but fail to add proper feature toggles or experiment flags---a standard practice in production mobile development for gradual rollout and A/B testing. \textbf{Missing Data Models} (22\%) occurs when agents fail to create or update data structures required by the PRD. \textbf{Missing Files} (11-15\%) represents cases where agents identify some but not all required files to modify. \textbf{Missing UI Components} (11-15\%) captures failures to implement specific UI elements like buttons, labels, or views. \textbf{Missing Required Methods} (9\%) reflects incomplete class implementations. While \textbf{Incomplete Multi-File Implementation} affects only 4-7\% of tasks, it disproportionately impacts complex features requiring coordination across 5+ files. The dominance of feature flag failures highlights a gap between agents' code generation capabilities and their understanding of production deployment practices.

\section{Discussion and Conclusion}

Our evaluation reveals a significant gap between current agent capabilities and the demands of industrial mobile development, with the best configurations achieving only a 12\% success rate. This shortfall, primarily driven by failures in cross-file reasoning and requirement understanding, underscores that autonomous software engineering remains an open challenge.

\xhdr{Implications}
For \textbf{practitioners}, our results suggest that agents should currently be viewed as ``copilots'' requiring human oversight rather than autonomous developers. The high variance in performance across agents for the same model (e.g., Cursor 12\% vs. OpenCode 2\% for Opus 4.5) highlights the critical role of agent scaffolding---practitioners should evaluate the complete system, not just the underlying model. The consistent gap between commercial and open-source agents suggests that engineering investment in tool integration and context management provides significant practical value. Furthermore, cost-effective models like GLM 4.6 can match the performance of expensive frontier models when paired with effective agent frameworks, offering a viable path for scalable adoption.
For \textbf{researchers}, the sharp performance drop on complex, multi-file tasks (18\% vs. 2\%) points to a need for better code context retrieval and graph-based reasoning. The 25\% failure rate due to requirement misunderstanding calls for improved grounding of natural language PRDs into code. Additionally, the under-utilization of visual designs suggests that future work must better integrate multi-modal signals into the coding loop.

\xhdr{Future Work}
We plan to expand SWE-Bench Mobile along several dimensions. First, we will add Android (Kotlin) tasks to enable cross-platform comparison and investigate whether agents exhibit consistent strengths and weaknesses across mobile ecosystems. Second, we will integrate simulator-based runtime evaluation to verify UI rendering, gesture handling, and state management---aspects that text-based diff inspection cannot capture. Third, we aim to evaluate additional open-source agents like OpenHands and SWE-Agent, and open-weight models like Qwen-Coder, to further broaden the benchmark's coverage. Finally, we plan to develop a public API for continuous evaluation, allowing agent providers to track their progress over time as both models and scaffolding improve.

\vspace{2mm}
In conclusion, SWE-Bench Mobile provides a rigorous testbed for the next generation of coding agents. While current performance is modest, the benchmark offers a clear roadmap for advancing agents from simple script generation to complex, industry-level software development.

\section{Related Work}
\subsection{Code Generation Benchmarks}

Early benchmarks for code generation focused on algorithmic problem-solving. \textbf{HumanEval}~\cite{humaneval} introduced 164 hand-crafted Python programming problems with unit tests, becoming a standard evaluation for code LLMs. \textbf{MBPP}~\cite{mbpp} expanded this with 974 crowd-sourced problems. While influential, these benchmarks test isolated function generation rather than realistic software engineering.

\textbf{SWE-Bench}~\cite{swebench} marked a significant advance by evaluating agents on real GitHub issues from popular Python repositories. Agents must understand issue descriptions, navigate codebases, and generate patches that pass existing tests. The benchmark has since evolved into a family of tasks, including \textbf{SWE-bench Multimodal}~\cite{swebench_multimodal}, which incorporates visual elements such as screenshots and diagrams to test visual software domains; \textbf{SWE-bench Multilingual}~\cite{swebench_multilingual,swesmith}, which expands evaluation to 9 programming languages beyond Python; and \textbf{SWE-bench Pro}~\cite{swebenchpro}, which introduces longer-horizon instances and includes proprietary/commercial codebases. Even with these extensions, many existing benchmarks still derive tasks from GitHub issue and pull-request artifacts, which more often emphasize bug fixing and localized improvements than new feature implementation from high-level specifications.

Other benchmarks target specific domains: \textbf{DS-1000}~\cite{ds1000} for data science, \textbf{ODEX}~\cite{odex} for open-domain execution, and \textbf{ClassEval}~\cite{classeval} for class-level generation. \textbf{DevBench}~\cite{devbench} evaluates repository-level coding but still focuses on Python.

SWE-Bench Mobile differs from these benchmarks in several key aspects: (1) multi-modal inputs including PRDs and Figma designs, (2) a large-scale production codebase (approx. 5GB), (3) mixed Swift/Objective-C target languages, and (4) feature implementation rather than bug fixing.

\subsection{Coding Agents}

The emergence of powerful LLMs has enabled a new generation of autonomous coding agents. These systems go beyond simple code completion to perform multi-step reasoning, tool use, and iterative refinement. 

\textbf{Commercial agents} include GitHub Copilot (Microsoft), Cursor (Anysphere), Claude Code (Anthropic), and Codex CLI (OpenAI). These agents integrate with development environments and can navigate codebases, run tests, and iterate on solutions.

\textbf{Open-source agents} have emerged as alternatives. OpenCode provides a terminal-based coding agent supporting multiple LLM backends. SWE-Agent~\cite{sweagent} introduces an agent-computer interface optimized for software engineering. AutoCodeRover~\cite{autocoderover} combines code search with LLM reasoning. Agentless~\cite{agentless} shows that simpler approaches without complex agent loops can be competitive. CodeAgent~\cite{codeagent} uses a repository-level code graph for navigation.

Our work provides a challenging benchmark for evaluating both commercial and open-source agents on industry-level tasks, revealing significant gaps in current capabilities and the importance of agent scaffolding.

\subsection{Multi-Modal Code Understanding}

Recent work has explored combining visual and textual information for code-related tasks. \textbf{Design2Code}~\cite{design2code} evaluates generating code from webpage screenshots. \textbf{Screenshot2Code} systems convert UI designs to implementation.

SWE-Bench Mobile extends this direction by incorporating Figma designs as part of the input specification, requiring agents to reason about visual layouts alongside textual requirements.

\subsection{Prompt Engineering for Code}

Prompt engineering significantly impacts LLM performance on coding tasks. Chain-of-thought prompting~\cite{cot} improves reasoning. Self-debugging~\cite{selfdebugging} enables iterative refinement. Structured prompts with role definitions and examples often outperform simple instructions.

Our ablation study (Section~\ref{sec:prompt_ablation}) systematically evaluates 12 prompt strategies, finding that ``Defensive Programming'' prompts emphasizing edge cases outperform both simple baselines and complex multi-step prompts.

\section*{Limitations}

\xhdr{Platform Scope}
SWE-Bench Mobile focuses on a single production iOS codebase from XiaoHongShu, which ensures depth and realism but limits generalization to other mobile platforms (Android, cross-platform frameworks like Flutter/React Native) and programming paradigms. The Swift/Objective-C mixed-language codebase, while representative of many large iOS projects, may not capture challenges unique to Kotlin-based Android development or cross-platform toolchains.

\xhdr{Evaluation Methodology}
Our evaluation uses text-based diff inspection rather than runtime execution, which means we validate structural correctness and architectural compliance but cannot detect issues that only manifest during runtime interactions, on specific devices, or under particular OS versions. Future work should integrate simulator-based testing to capture dynamic behaviors such as UI rendering, memory management, and concurrency issues.

\xhdr{Prompt and Model Coverage}
Our prompt ablation study covers one agent-model configuration (Claude Code + GLM 4.6) and 12 prompt variants. While this provides insights into prompt sensitivity, different models may respond differently to these strategies. Additionally, API costs reported are based on pricing at experiment time and may vary with different prompting strategies or model updates.

\xhdr{Benchmark Scale}
The benchmark's 50 tasks, while derived from real product development, represent a snapshot of mobile development challenges and may not cover all possible feature types (e.g., real-time communication, payment integration, accessibility features) or edge cases encountered in production. We plan to continuously expand the task set to improve coverage.

\section*{Ethics Statement}

The tasks and codebase in SWE-Bench Mobile are derived from XiaoHongShu Inc. with explicit permission for research use. The codebase snapshot excludes sensitive credentials and business logic. Human validation was performed by the authors and XiaoHongShu engineers; no crowdworkers were employed.

Our work evaluates AI agents for software engineering tasks. Current performance (12\% task success rate) indicates that human oversight remains essential. We view these agents as assistive tools rather than replacements for human developers. Practitioners should use comprehensive testing and code review when deploying AI-generated code, as emphasized by our benchmark's evaluation approach.

\clearpage
\onecolumn

\bibliographystyle{ACM-Reference-Format}
\bibliography{custom}


\begin{thebibliography}{19}


\ifx \showCODEN    \undefined \def \showCODEN     #1{\unskip}     \fi
\ifx \showISBNx    \undefined \def \showISBNx     #1{\unskip}     \fi
\ifx \showISBNxiii \undefined \def \showISBNxiii  #1{\unskip}     \fi
\ifx \showISSN     \undefined \def \showISSN      #1{\unskip}     \fi
\ifx \showLCCN     \undefined \def \showLCCN      #1{\unskip}     \fi
\ifx \shownote     \undefined \def \shownote      #1{#1}          \fi
\ifx \showarticletitle \undefined \def \showarticletitle #1{#1}   \fi
\ifx \showURL      \undefined \def \showURL       {\relax}        \fi
\providecommand\bibfield[2]{#2}
\providecommand\bibinfo[2]{#2}
\providecommand\natexlab[1]{#1}
\providecommand\showeprint[2][]{arXiv:#2}

\bibitem[Atlassian(2024)]%
        {atlassian_prd}
\bibfield{author}{\bibinfo{person}{Atlassian}.}
  \bibinfo{year}{2024}\natexlab{}.
\newblock \bibinfo{title}{How to Write a Product Requirements Document (PRD)}.
\newblock
  \bibinfo{howpublished}{\url{https://www.atlassian.com/agile/product-management/requirements}}.
\newblock


\bibitem[Austin et~al\mbox{.}(2021)]%
        {mbpp}
\bibfield{author}{\bibinfo{person}{Jacob Austin}, \bibinfo{person}{Augustus
  Odena}, \bibinfo{person}{Maxwell Nye}, \bibinfo{person}{Maarten Bosma},
  \bibinfo{person}{Henryk Michalewski}, \bibinfo{person}{David Dohan},
  \bibinfo{person}{Ellen Jiang}, \bibinfo{person}{Carrie Cai},
  \bibinfo{person}{Michael Terry}, \bibinfo{person}{Quoc Le}, {et~al\mbox{.}}}
  \bibinfo{year}{2021}\natexlab{}.
\newblock \showarticletitle{Program Synthesis with Large Language Models}.
\newblock \bibinfo{journal}{\emph{arXiv preprint arXiv:2108.07732}}
  (\bibinfo{year}{2021}).
\newblock


\bibitem[Chen et~al\mbox{.}(2021)]%
        {humaneval}
\bibfield{author}{\bibinfo{person}{Mark Chen}, \bibinfo{person}{Jerry Tworek},
  \bibinfo{person}{Heewoo Jun}, \bibinfo{person}{Qiming Yuan},
  \bibinfo{person}{Henrique Ponde de~Oliveira Pinto}, \bibinfo{person}{Jared
  Kaplan}, \bibinfo{person}{Harri Edwards}, \bibinfo{person}{Yuri Burda},
  \bibinfo{person}{Nicholas Joseph}, \bibinfo{person}{Greg Brockman},
  {et~al\mbox{.}}} \bibinfo{year}{2021}\natexlab{}.
\newblock \showarticletitle{Evaluating Large Language Models Trained on Code}.
\newblock \bibinfo{journal}{\emph{arXiv preprint arXiv:2107.03374}}
  (\bibinfo{year}{2021}).
\newblock


\bibitem[Chen et~al\mbox{.}(2023)]%
        {selfdebugging}
\bibfield{author}{\bibinfo{person}{Xinyun Chen}, \bibinfo{person}{Maxwell Lin},
  \bibinfo{person}{Nathanael Sch{\"a}rli}, {and} \bibinfo{person}{Denny Zhou}.}
  \bibinfo{year}{2023}\natexlab{}.
\newblock \showarticletitle{Teaching Large Language Models to Self-Debug}.
\newblock \bibinfo{journal}{\emph{arXiv preprint arXiv:2304.05128}}
  (\bibinfo{year}{2023}).
\newblock


\bibitem[Deng et~al\mbox{.}(2025)]%
        {swebenchpro}
\bibfield{author}{\bibinfo{person}{Xiang Deng}, \bibinfo{person}{Jeff Da},
  \bibinfo{person}{Edwin Pan}, \bibinfo{person}{Yannis~Yiming He},
  \bibinfo{person}{Charles Ide}, \bibinfo{person}{Kanak Garg},
  \bibinfo{person}{Niklas Lauffer}, \bibinfo{person}{Andrew Park},
  \bibinfo{person}{Nitin Pasari}, \bibinfo{person}{Chetan Rane},
  {et~al\mbox{.}}} \bibinfo{year}{2025}\natexlab{}.
\newblock \showarticletitle{SWE-Bench Pro: Can AI Agents Solve Long-Horizon
  Software Engineering Tasks?}
\newblock \bibinfo{journal}{\emph{arXiv preprint arXiv:2509.16941}}
  (\bibinfo{year}{2025}).
\newblock


\bibitem[Du et~al\mbox{.}(2024)]%
        {classeval}
\bibfield{author}{\bibinfo{person}{Xueying Du}, \bibinfo{person}{Mingwei Liu},
  \bibinfo{person}{Kaixin Wang}, \bibinfo{person}{Hanlin Wang},
  \bibinfo{person}{Junwei Liu}, \bibinfo{person}{Yixuan Chen},
  \bibinfo{person}{Jiayi Feng}, \bibinfo{person}{Chaofeng Sha},
  \bibinfo{person}{Xin Peng}, {and} \bibinfo{person}{Yiling Lou}.}
  \bibinfo{year}{2024}\natexlab{}.
\newblock \showarticletitle{ClassEval: A Manually-Crafted Benchmark for
  Evaluating LLMs on Class-level Code Generation}. In
  \bibinfo{booktitle}{\emph{International Conference on Machine Learning}}.
\newblock


\bibitem[Jimenez et~al\mbox{.}(2024)]%
        {swebench}
\bibfield{author}{\bibinfo{person}{Carlos~E Jimenez}, \bibinfo{person}{John
  Yang}, \bibinfo{person}{Alexander Wettig}, \bibinfo{person}{Shunyu Yao},
  \bibinfo{person}{Kexin Pei}, \bibinfo{person}{Ofir Press}, {and}
  \bibinfo{person}{Karthik~R Narasimhan}.} \bibinfo{year}{2024}\natexlab{}.
\newblock \showarticletitle{SWE-bench: Can Language Models Resolve Real-world
  GitHub Issues?}. In \bibinfo{booktitle}{\emph{The Twelfth International
  Conference on Learning Representations}}.
\newblock


\bibitem[Lai et~al\mbox{.}(2023)]%
        {ds1000}
\bibfield{author}{\bibinfo{person}{Yuhang Lai}, \bibinfo{person}{Chengxi Li},
  \bibinfo{person}{Yiming Wang}, \bibinfo{person}{Tianyi Zhang},
  \bibinfo{person}{Ruiqi Zhong}, \bibinfo{person}{Luke Zettlemoyer},
  \bibinfo{person}{Wen-tau Yih}, \bibinfo{person}{Daniel Fried},
  \bibinfo{person}{Sida Wang}, {and} \bibinfo{person}{Tao Yu}.}
  \bibinfo{year}{2023}\natexlab{}.
\newblock \showarticletitle{DS-1000: A Natural and Reliable Benchmark for Data
  Science Code Generation}. In \bibinfo{booktitle}{\emph{International
  Conference on Machine Learning}}.
\newblock


\bibitem[Li et~al\mbox{.}(2024)]%
        {devbench}
\bibfield{author}{\bibinfo{person}{Bowen Li}, \bibinfo{person}{Wenhan Wu},
  \bibinfo{person}{Ziwei Tang}, \bibinfo{person}{Lin Shi},
  \bibinfo{person}{John Yang}, \bibinfo{person}{Jinyang Li},
  \bibinfo{person}{Shunyu Yao}, \bibinfo{person}{Chen Xiong}, {and}
  \bibinfo{person}{Karthik Narasimhan}.} \bibinfo{year}{2024}\natexlab{}.
\newblock \showarticletitle{DevBench: A Comprehensive Benchmark for Software
  Development}.
\newblock \bibinfo{journal}{\emph{arXiv preprint arXiv:2403.08604}}
  (\bibinfo{year}{2024}).
\newblock


\bibitem[Si et~al\mbox{.}(2024)]%
        {design2code}
\bibfield{author}{\bibinfo{person}{Chenglei Si}, \bibinfo{person}{Yanzhe Li},
  \bibinfo{person}{Zhengyuan Jiang}, \bibinfo{person}{Xinyang Liu},
  \bibinfo{person}{Zheng Lu}, \bibinfo{person}{Yuqing Jiang},
  \bibinfo{person}{Yong Liu}, \bibinfo{person}{Yu Wang}, \bibinfo{person}{Yujiu
  Yuan}, \bibinfo{person}{Lydia Liu}, {et~al\mbox{.}}}
  \bibinfo{year}{2024}\natexlab{}.
\newblock \showarticletitle{Design2Code: How Far Are We From Automating
  Front-End Engineering?}
\newblock \bibinfo{journal}{\emph{arXiv preprint arXiv:2403.03163}}
  (\bibinfo{year}{2024}).
\newblock


\bibitem[Wang et~al\mbox{.}(2022)]%
        {odex}
\bibfield{author}{\bibinfo{person}{Zhiruo Wang}, \bibinfo{person}{Shuyan Zhou},
  \bibinfo{person}{Daniel Fried}, {and} \bibinfo{person}{Graham Neubig}.}
  \bibinfo{year}{2022}\natexlab{}.
\newblock \showarticletitle{Execution-Based Evaluation for Open-Domain Code
  Generation}.
\newblock \bibinfo{journal}{\emph{arXiv preprint arXiv:2212.10481}}
  (\bibinfo{year}{2022}).
\newblock


\bibitem[Wei et~al\mbox{.}(2022)]%
        {cot}
\bibfield{author}{\bibinfo{person}{Jason Wei}, \bibinfo{person}{Xuezhi Wang},
  \bibinfo{person}{Dale Schuurmans}, \bibinfo{person}{Maarten Bosma},
  \bibinfo{person}{Fei Xia}, \bibinfo{person}{Ed Chi}, \bibinfo{person}{Quoc~V
  Le}, \bibinfo{person}{Denny Zhou}, {et~al\mbox{.}}}
  \bibinfo{year}{2022}\natexlab{}.
\newblock \showarticletitle{Chain-of-Thought Prompting Elicits Reasoning in
  Large Language Models}.
\newblock \bibinfo{journal}{\emph{Advances in Neural Information Processing
  Systems}}  \bibinfo{volume}{35} (\bibinfo{year}{2022}),
  \bibinfo{pages}{24824--24837}.
\newblock


\bibitem[Xia et~al\mbox{.}(2024)]%
        {agentless}
\bibfield{author}{\bibinfo{person}{Chunqiu~Steven Xia}, \bibinfo{person}{Yinlin
  Deng}, \bibinfo{person}{Soren Dunn}, {and} \bibinfo{person}{Lingming Zhang}.}
  \bibinfo{year}{2024}\natexlab{}.
\newblock \showarticletitle{Agentless: Demystifying LLM-based Software
  Engineering Agents}.
\newblock \bibinfo{journal}{\emph{arXiv preprint arXiv:2407.01489}}
  (\bibinfo{year}{2024}).
\newblock


\bibitem[Yang et~al\mbox{.}(2024)]%
        {sweagent}
\bibfield{author}{\bibinfo{person}{John Yang}, \bibinfo{person}{Carlos~E
  Jimenez}, \bibinfo{person}{Alexander Wettig}, \bibinfo{person}{Kilian
  Lieret}, \bibinfo{person}{Shunyu Yao}, \bibinfo{person}{Karthik Narasimhan},
  {and} \bibinfo{person}{Ofir Press}.} \bibinfo{year}{2024}\natexlab{}.
\newblock \showarticletitle{SWE-agent: Agent-Computer Interfaces Enable
  Automated Software Engineering}.
\newblock \bibinfo{journal}{\emph{arXiv preprint arXiv:2405.15793}}
  (\bibinfo{year}{2024}).
\newblock


\bibitem[Yang et~al\mbox{.}(2025a)]%
        {swebench_multimodal}
\bibfield{author}{\bibinfo{person}{John Yang}, \bibinfo{person}{Carlos~E
  Jimenez}, \bibinfo{person}{Alex~L Zhang}, \bibinfo{person}{Kilian Lieret},
  \bibinfo{person}{Joyce Yang}, \bibinfo{person}{Xindi Wu},
  \bibinfo{person}{Ofir Press}, \bibinfo{person}{Niklas Muennighoff},
  \bibinfo{person}{Gabriel Synnaeve}, \bibinfo{person}{Karthik~R Narasimhan},
  {et~al\mbox{.}}} \bibinfo{year}{2025}\natexlab{a}.
\newblock \showarticletitle{SWE-bench Multimodal: Do AI Systems Generalize to
  Visual Software Domains?}. In \bibinfo{booktitle}{\emph{The Thirteenth
  International Conference on Learning Representations}}.
\newblock


\bibitem[Yang et~al\mbox{.}(2025b)]%
        {swesmith}
\bibfield{author}{\bibinfo{person}{John Yang}, \bibinfo{person}{Kilian Lieret},
  \bibinfo{person}{Carlos~E Jimenez}, \bibinfo{person}{Alexander Wettig},
  \bibinfo{person}{Kabir Khandpur}, \bibinfo{person}{Yanzhe Zhang},
  \bibinfo{person}{Binyuan Hui}, \bibinfo{person}{Ofir Press},
  \bibinfo{person}{Ludwig Schmidt}, {and} \bibinfo{person}{Diyi Yang}.}
  \bibinfo{year}{2025}\natexlab{b}.
\newblock \showarticletitle{SWE-smith: Scaling Data for Software Engineering
  Agents}.
\newblock \bibinfo{journal}{\emph{arXiv preprint arXiv:2504.21798}}
  (\bibinfo{year}{2025}).
\newblock


\bibitem[Zan et~al\mbox{.}(2025)]%
        {swebench_multilingual}
\bibfield{author}{\bibinfo{person}{Daoguang Zan}, \bibinfo{person}{Zhirong
  Huang}, \bibinfo{person}{Wei Liu}, \bibinfo{person}{Hanwu Chen},
  \bibinfo{person}{Linhao Zhang}, \bibinfo{person}{Shulin Xin},
  \bibinfo{person}{Lu Chen}, \bibinfo{person}{Qi Liu},
  \bibinfo{person}{Xiaojian Zhong}, \bibinfo{person}{Aoyan Li},
  \bibinfo{person}{Siyao Liu}, \bibinfo{person}{Yongsheng Xiao},
  \bibinfo{person}{Liangqiang Chen}, \bibinfo{person}{Yuyu Zhang},
  \bibinfo{person}{Jing Su}, \bibinfo{person}{Tianyu Liu}, \bibinfo{person}{Rui
  Long}, \bibinfo{person}{Kai Shen}, {and} \bibinfo{person}{Liang Xiang}.}
  \bibinfo{year}{2025}\natexlab{}.
\newblock \bibinfo{title}{Multi-SWE-bench: A Multilingual Benchmark for Issue
  Resolving}.
\newblock
\showeprint[arxiv]{2504.02605}~[cs.SE]
\urldef\tempurl%
\url{https://arxiv.org/abs/2504.02605}
\showURL{%
\tempurl}


\bibitem[Zhang et~al\mbox{.}(2024a)]%
        {codeagent}
\bibfield{author}{\bibinfo{person}{Kechi Zhang}, \bibinfo{person}{Jia Li},
  \bibinfo{person}{Ge Li}, \bibinfo{person}{Xianjie Shi}, {and}
  \bibinfo{person}{Zhi Jin}.} \bibinfo{year}{2024}\natexlab{a}.
\newblock \showarticletitle{CodeAgent: Enhancing Code Generation with
  Tool-Integrated Agent Systems for Real-World Repo-level Coding Challenges}.
\newblock \bibinfo{journal}{\emph{arXiv preprint arXiv:2401.07339}}
  (\bibinfo{year}{2024}).
\newblock


\bibitem[Zhang et~al\mbox{.}(2024b)]%
        {autocoderover}
\bibfield{author}{\bibinfo{person}{Yuntong Zhang}, \bibinfo{person}{Haifeng
  Ruan}, \bibinfo{person}{Zhiyu Fan}, {and} \bibinfo{person}{Abhik
  Roychoudhury}.} \bibinfo{year}{2024}\natexlab{b}.
\newblock \showarticletitle{AutoCodeRover: Autonomous Program Improvement}.
\newblock \bibinfo{journal}{\emph{arXiv preprint arXiv:2404.05427}}
  (\bibinfo{year}{2024}).
\newblock


\end{thebibliography}

\clearpage
\appendix

\definecolor{prdblue}{RGB}{235,245,255}
\definecolor{prdborder}{RGB}{59,130,246}
\definecolor{codegray}{RGB}{249,250,251}
\definecolor{codeborder}{RGB}{209,213,219}
\definecolor{successgreen}{RGB}{236,253,245}
\definecolor{successborder}{RGB}{34,197,94}
\definecolor{warningred}{RGB}{254,242,242}
\definecolor{warningborder}{RGB}{239,68,68}
\definecolor{accentpurple}{RGB}{245,243,255}
\definecolor{accentborder}{RGB}{139,92,246}
\definecolor{neutralgray}{RGB}{107,114,128}
\definecolor{darktext}{RGB}{31,41,55}

\newtcolorbox{prdbox}[1][]{
    enhanced,
    breakable,
    colback=prdblue,
    colframe=prdborder,
    boxrule=1pt,
    arc=4pt,
    left=10pt,
    right=10pt,
    top=10pt,
    bottom=10pt,
    fonttitle=\bfseries\sffamily,
    coltitle=white,
    attach boxed title to top left={yshift=-2mm, xshift=4mm},
    boxed title style={colback=prdborder, arc=3pt, boxrule=0pt},
    shadow={1mm}{-1mm}{0mm}{black!10},
    #1
}

\newtcolorbox{codebox}[1][]{
    enhanced,
    breakable,
    colback=codegray,
    colframe=codeborder,
    boxrule=0.8pt,
    arc=3pt,
    left=8pt,
    right=8pt,
    top=6pt,
    bottom=6pt,
    fontupper=\ttfamily\small,
    #1
}

\newtcolorbox{successbox}[1][]{
    enhanced,
    breakable,
    colback=successgreen,
    colframe=successborder,
    boxrule=1pt,
    arc=4pt,
    left=10pt,
    right=10pt,
    top=10pt,
    bottom=10pt,
    fonttitle=\bfseries\sffamily,
    coltitle=white,
    attach boxed title to top left={yshift=-2mm, xshift=4mm},
    boxed title style={colback=successborder, arc=3pt, boxrule=0pt},
    #1
}

\newtcolorbox{warningbox}[1][]{
    enhanced,
    breakable,
    colback=warningred,
    colframe=warningborder,
    boxrule=1pt,
    arc=4pt,
    left=10pt,
    right=10pt,
    top=10pt,
    bottom=10pt,
    fonttitle=\bfseries\sffamily,
    coltitle=white,
    attach boxed title to top left={yshift=-2mm, xshift=4mm},
    boxed title style={colback=warningborder, arc=3pt, boxrule=0pt},
    #1
}

\newcommand{\taskheader}[3]{%
    \noindent
    \fcolorbox{codeborder}{codegray}{%
        \parbox{\dimexpr\linewidth-2\fboxsep-2\fboxrule}{%
            \small
            \textcolor{neutralgray}{\textsf{Difficulty:}} \textbf{#1} \hfill
            \textcolor{neutralgray}{\textsf{Files to Modify:}} \textbf{#2} \hfill
            \textcolor{neutralgray}{\textsf{Test Cases:}} \textbf{#3}%
        }%
    }%
    \vspace{4mm}
}

\newcommand{\sectiondivider}{%
    \vspace{4mm}
    \noindent\textcolor{prdborder}{\rule{\linewidth}{0.5pt}}
    \vspace{4mm}
}

\section{Task Examples}
\label{app:task_example}

We present two representative tasks from SWE-Bench Mobile to illustrate the benchmark format. Each task includes a Product Requirement Document (PRD) with design specifications, translated from the original Chinese used by the development team.

\sectiondivider

\subsection{Task 003: Custom Emoji Limit Adjustment}
\label{app:task003}

\taskheader{Easy}{3}{5}

\begin{prdbox}[title={\small Adjust Custom Emoji Collection Limit}]
\textbf{\textsf{Background}}
\vspace{1.5mm}

The current custom emoji (saved stickers) limit is hardcoded to 300 on the client side. As user demand grows, we need to increase this limit to better serve our users.

\vspace{2mm}
\textbf{\textsf{Requirements}}
\vspace{1.5mm}

\begin{enumerate}[leftmargin=1.5em, topsep=3pt, itemsep=2pt, parsep=0pt]
    \item \textbf{Increase limit:} Change from 300 to 999
    \item \textbf{Update UI prompts:} Adjust warning messages to reflect new limit
    \item \textbf{Server-driven config:} Remove hardcoded values; future changes should not require app updates
    \item \textbf{Comprehensive coverage:} Apply to all emoji-saving scenarios (chat, comments, etc.)
\end{enumerate}

\vspace{2mm}
\textbf{\textsf{Competitor Analysis}}
\vspace{2mm}

\begin{center}
\renewcommand{\arraystretch}{1.3}
\begin{tabular}{@{}lc@{}}
\toprule
\textbf{App} & \textbf{Emoji Limit} \\
\midrule
WeChat & 999 \\
Douyin (TikTok) & 599 \\
Kuaishou & 158 \\
\bottomrule
\end{tabular}
\end{center}
\end{prdbox}

\vspace{4mm}
\noindent\textbf{\textsf{Design Mockups}} \hspace{2mm} See Figure~\ref{fig:task003_mockups} for the original design specifications provided to developers.

\vspace{4mm}
\noindent\textbf{\textsf{Evaluation Criteria}}
\vspace{2mm}

\begin{itemize}[leftmargin=1.5em, topsep=3pt, itemsep=2pt, parsep=0pt]
    \item[\textcolor{successborder}{\checkmark}] Hardcoded limit (300) removed or increased to $\geq$450
    \item[\textcolor{successborder}{\checkmark}] New limit (999) properly configured
    \item[\textcolor{successborder}{\checkmark}] Server-driven configuration implemented
    \item[\textcolor{successborder}{\checkmark}] Changes applied across multiple files
    \item[\textcolor{successborder}{\checkmark}] Non-empty, meaningful code changes
\end{itemize}

\sectiondivider

\subsection{Task 007: Card Message Click Decoupling}
\label{app:task007}

\taskheader{Medium}{5}{5}

\begin{prdbox}[title={\small iOS Card Reference Click Decoupling}]
\textbf{\textsf{Background}}
\vspace{1.5mm}

Card messages have been added to the app. While most iOS code is decoupled from the messaging module, the click logic for card message references remains coupled in \texttt{AppChatBaseViewController}. This task decouples the click handling for better maintainability.

\vspace{2mm}
\textbf{\textsf{Architecture Design}}
\vspace{1.5mm}

Abstract click logic to \texttt{CardRefBaseProvider}. The view controller should find the concrete implementation based on card type, following the provider pattern.

\vspace{2mm}
\textbf{\textsf{Implementation Sketch}}
\vspace{2mm}

\begin{codebox}
\begin{lstlisting}[language=Swift, basicstyle=\ttfamily\small, frame=none, xleftmargin=0pt, xrightmargin=0pt]
@objc(AppRefMessageDataService)
public class AppRefMessageDataService: NSObject {
    var chatType: String?
    var chatId: String?
    var senderId: String?
    var messageId: String?
}
\end{lstlisting}
\end{codebox}

\vspace{3mm}
\noindent\textbf{\textsf{Impact Scope:}} Shopping card, Advertisement card
\end{prdbox}

\vspace{4mm}
\noindent\textbf{\textsf{Evaluation Criteria}}
\vspace{2mm}

\begin{itemize}[leftmargin=1.5em, topsep=3pt, itemsep=2pt, parsep=0pt]
    \item[\textcolor{successborder}{\checkmark}] New \texttt{AppRefMessageDataService} class created
    \item[\textcolor{successborder}{\checkmark}] Click handling moved out of \texttt{AppChatBaseViewController}
    \item[\textcolor{successborder}{\checkmark}] Provider pattern correctly implemented
    \item[\textcolor{successborder}{\checkmark}] Shopping and advertisement card handling works
    \item[\textcolor{successborder}{\checkmark}] No regression in existing functionality
\end{itemize}

\section{Complete Experimental Results}
\label{app:detailed_results}

\begin{table}[t]
\centering
\caption{Complete evaluation results on SWE-Bench Mobile. \textbf{Task Success} measures the percentage of tasks where all test cases pass (out of 50 tasks). \textbf{Test Pass} measures the percentage of individual test cases passed (out of 449 tests). Best results per agent in \textbf{bold}.}
\label{tab:full_results}
\small
\setlength{\tabcolsep}{8pt}
\renewcommand{\arraystretch}{1.25}
\begin{tabular}{@{}ll cc cc@{}}
\toprule
\textbf{Agent} & \textbf{Model} & \textbf{Task Success (\%)} & \textbf{Test Pass (\%)} & \textbf{Cost (\$/task)} & \textbf{Time (min)} \\
\midrule
\multirow{5}{*}{\textsc{Cursor}} 
    & Claude Opus 4.5 & \textbf{12.0} & \textbf{28.1} & 3.50 & 15.0 \\
    & Claude Sonnet 4.5 & \textbf{12.0} & 26.7 & 2.00 & 14.2 \\
    & GPT-5.2 & 8.0 & 27.4 & 1.80 & 20.0 \\
    & Gemini 3 Pro & 6.0 & 23.2 & 1.00 & 12.5 \\
    & GPT-5.1 & 2.0 & 19.6 & 1.10 & 14.2 \\
\midrule
\multirow{5}{*}{\textsc{Codex}}
    & GLM-4.6 & \textbf{12.0} & 19.6 & 1.30 & 13.3 \\
    & Claude Sonnet 4.5 & 10.0 & \textbf{28.1} & 2.50 & 12.5 \\
    & GPT-5 & 10.0 & 21.4 & 1.50 & 10.0 \\
    & Claude Opus 4.5 & 4.0 & 20.7 & 3.50 & 14.2 \\
    & GPT-5.1 & 0.0 & 7.1 & 1.00 & 13.3 \\
\midrule
\multirow{4}{*}{\textsc{Claude Code}}
    & GLM-4.6 & \textbf{10.0} & \textbf{26.7} & 1.30 & 11.7 \\
    & Claude Sonnet 4.5 & 10.0 & 24.7 & 2.00 & 13.3 \\
    & Claude Opus 4.5 & 8.0 & 21.8 & 4.00 & 15.0 \\
    & Claude Haiku & 8.0 & 18.3 & 0.50 & 8.3 \\
\midrule
\multirow{8}{*}{\textsc{OpenCode}}
    & GLM-4.6 & \textbf{8.0} & \textbf{17.8} & 0.13 & 32.5 \\
    & GPT-5.1 & 6.0 & 7.1 & 0.02 & 2.0 \\
    & Claude Sonnet 4.5 & 4.0 & 14.7 & 3.50 & 11.1 \\
    & GLM-4.7 & 4.0 & 14.3 & 0.49 & 52.1 \\
    & Gemini 3 Pro & 4.0 & 13.4 & 0.03 & 8.9 \\
    & GPT-5.2 & 4.0 & 12.0 & 0.04 & 10.9 \\
    & Claude Opus 4.5 & 2.0 & 12.0 & 9.33 & 8.2 \\
    & GPT-5 & 2.0 & 12.0 & 0.18 & 9.8 \\
\bottomrule
\end{tabular}
\end{table}

\subsection{Cross-Agent Model Comparison}

Table~\ref{tab:model_comparison} reveals that the same model can perform very differently across agents, highlighting the importance of agent design.

\begin{table}[t]
\centering
\caption{Same model, different agents: Task Success Rate (\%). The gap between best and worst agent can be as large as 6$\times$.}
\label{tab:model_comparison}
\small
\setlength{\tabcolsep}{5pt}
\renewcommand{\arraystretch}{1.25}
\begin{tabular}{@{}l cccc c@{}}
\toprule
\textbf{Model} & \textbf{Cursor} & \textbf{Codex} & \textbf{CC} & \textbf{OpenCode} & \textbf{Gap} \\
\midrule
Opus 4.5 & \textbf{12} & 4 & 8 & 2 & 6$\times$ \\
Sonnet 4.5 & \textbf{12} & 10 & 10 & 4 & 3$\times$ \\
GLM-4.6 & --- & \textbf{12} & 10 & 8 & 1.5$\times$ \\
GPT-5.1 & 2 & 0 & --- & \textbf{6} & $\infty$ \\
\bottomrule
\end{tabular}
\end{table}

\clearpage
\section{Prompt Templates}
\label{app:prompt_ablation}

We designed 12 prompt variants for the ablation study. Below we present the key prompts. All prompts share a common structure: role definition, task description, and output format. The differentiating factor is the \textit{emphasis} placed on different aspects.

\subsection{Best Prompt: Defensive Programming}

\begin{successbox}[title={\small P10: Defensive Programming (Best)}]
\itshape
``You are a senior iOS engineer known for writing robust, production-ready code. Implement the feature with a focus on \textbf{defensive programming} and edge case handling.

\vspace{3mm}
Don't just implement the happy path. Think about everything that could go wrong:

\vspace{1mm}
\begin{itemize}[leftmargin=1.5em, topsep=2pt, itemsep=1pt, parsep=0pt]
    \item Empty data, nil values, invalid formats
    \item Very long/short text, different screen sizes
    \item Slow network, timeouts, concurrent operations
    \item First-time user, offline mode, low memory
\end{itemize}

\vspace{3mm}
Your code should handle all of this gracefully without crashing.''
\end{successbox}

\subsection{Baseline Prompt}

\begin{prdbox}[title={\small P1: Baseline}]
\itshape
``You are an iOS developer. Read the PRD carefully and implement the required changes. Generate a unified diff patch that can be applied to the codebase.''
\end{prdbox}

\subsection{Worst Performing Prompts}

\begin{warningbox}[title={\small P12: Comprehensive (Worst)}]
\itshape
``You are a senior iOS engineer. Before implementing:

\vspace{1mm}
\begin{enumerate}[leftmargin=1.5em, topsep=2pt, itemsep=1pt, parsep=0pt]
    \item Analyze the PRD thoroughly
    \item Identify all affected files
    \item Plan your implementation strategy
    \item Consider edge cases
    \item Review the Figma design
    \item Check for existing patterns
    \item Implement with tests in mind
    \item Validate against requirements
\end{enumerate}

\vspace{2mm}
Generate a complete, production-ready patch.''
\end{warningbox}

\vspace{4mm}
\noindent\fcolorbox{warningborder}{warningred}{\parbox{0.95\linewidth}{%
\small\textbf{\textsf{Why Comprehensive Failed:}} The overly detailed checklist appears to overwhelm the model, causing it to focus on process rather than actual implementation. Simpler, focused prompts consistently outperform complex ones.
}}

\subsection{Other Notable Prompts}

\vspace{3mm}
\begin{description}[leftmargin=0pt, labelindent=0pt, itemsep=4pt]
    \item[\textsf{P7: Chain of Thought}] Asks the model to ``think step by step'' before coding. Achieved 10\% Task Success but lower Test Pass Rate (21.8\%) than Defensive Programming.

    \item[\textsf{P9: Figma Emphasis}] Emphasizes matching the Figma design exactly. Surprisingly underperformed (8\% Task Success), possibly because many tasks don't require UI changes.

    \item[\textsf{P11: Test Driven}] Asks the model to ``think about what tests would verify your implementation.'' Achieved only 6\% Task Success despite the intuitive appeal of test-driven thinking.
\end{description}

\section{Dataset Statistics}
\label{app:dataset_stats}

We provide detailed statistics of the SWE-Bench Mobile dataset in Table~\ref{tab:dataset_stats}. The benchmark consists of 50 tasks with varying levels of complexity, involving multi-modal inputs (PRDs and Figma designs) and a large-scale production codebase. The tasks are designed to cover a wide range of mobile development scenarios, ensuring a comprehensive evaluation of agent capabilities.

\begin{table}[t]
\centering
\caption{SWE-Bench Mobile dataset statistics.}
\label{tab:dataset_stats}
\small
\setlength{\tabcolsep}{12pt}
\renewcommand{\arraystretch}{1.3}
\begin{tabular}{@{}lr@{}}
\toprule
\textbf{Metric} & \textbf{Value} \\
\midrule
\rowcolor{prdblue}
\multicolumn{2}{@{}l}{\textit{\textsf{Task Composition}}} \\
\quad Total Tasks & 50 \\
\quad Tasks with Figma Design & 35 (70\%) \\
\quad Tasks with Reference Images & 46 (92\%) \\
\midrule
\rowcolor{prdblue}
\multicolumn{2}{@{}l}{\textit{\textsf{Task Complexity}}} \\
\quad Avg. PRD Length (words) & 450 \\
\quad Avg. Test Cases per Task & 9.1 \\
\quad Total Test Cases & 449 \\
\quad Avg. Files to Modify & 4.2 \\
\midrule
\rowcolor{prdblue}
\multicolumn{2}{@{}l}{\textit{\textsf{Codebase}}} \\
\quad Programming Language & Swift/Objective-C (iOS) \\
\quad Codebase Size & $\sim$500K LoC \\
\bottomrule
\end{tabular}
\end{table}

\section{Reproducibility}
\label{app:reproducibility}

\xhdr{Environment}
All experiments were conducted on macOS 14.x with:

\vspace{2mm}
\begin{itemize}[leftmargin=1.5em, topsep=3pt, itemsep=3pt, parsep=0pt]
    \item \textsc{Cursor}: v2.3 with Agent mode enabled
    \item \textsc{Codex}: OpenAI Codex CLI v0.77.0
    \item \textsc{Claude Code}: Anthropic Claude Code CLI v2.1.37
    \item \textsc{OpenCode}: v1.1.44 (open-source coding agent)
\end{itemize}

\xhdr{Model API Configuration}

For reproducibility, we specify the exact API endpoints and configurations used:

\vspace{2mm}
\begin{itemize}[leftmargin=1.5em, topsep=3pt, itemsep=3pt, parsep=0pt]
    \item \textbf{GPT Models (GPT 5, 5.1, 5.2)}: Accessed via Microsoft Azure OpenAI API with default temperature and top-p settings
    \item \textbf{Claude Models (Opus 4.5, Sonnet 4.5, Haiku)}: Accessed via Google Vertex AI API for Anthropic models
    \item \textbf{Gemini 3 Pro}: Accessed via Google Vertex AI API with standard configuration
    \item \textbf{GLM Models (GLM 4.6, 4.7)}: Used GLM Coding Plan with default agent scaffolding
\end{itemize}

\xhdr{Multi-Modal Input Handling}

To handle Figma designs and reference images, we configured Model Context Protocol (MCP) integrations:

\vspace{2mm}
\begin{itemize}[leftmargin=1.5em, topsep=3pt, itemsep=3pt, parsep=0pt]
    \item \textbf{Vision-capable models (GPT, Claude, Gemini)}: Used official Figma MCP to directly access design specifications
    \item \textbf{GLM Models}: Since GLM 4.6 is not a native vision model, we used the official GLM Vision MCP to process images and Figma designs, converting visual inputs into structured descriptions for the text-only model
\end{itemize}

\xhdr{Evaluation Pipeline}

\vspace{2mm}
\begin{enumerate}[leftmargin=1.5em, topsep=3pt, itemsep=3pt, parsep=0pt]
    \item Load generated patch file as text
    \item Run task-specific pytest test suite (tests inspect the patch diff text using pattern matching and structural analysis)
    \item Record pass/fail status for each test case
    \item Aggregate results across all 50 tasks
\end{enumerate}

\xhdr{Availability and Hosted Evaluation}
The SWE-Bench Mobile benchmark is derived from a proprietary production codebase with permission from XiaoHongShu Inc. Due to the confidential nature of the source code and product requirements, the full dataset cannot be publicly released. We view this constraint as a feature rather than a limitation: by keeping the test set private, we \textit{eliminate the risk of data contamination}---a well-known issue with public benchmarks where test instances may leak into LLM training corpora~\cite{swebench}.

SWE-Bench Mobile is designed as a \textit{standardized evaluation platform for coding agent providers and foundation model vendors}. We host a public leaderboard at \url{https://swebenchmobile.com} where agent companies (e.g., Cursor, Codex, Claude Code) and model providers (e.g., OpenAI, Anthropic, Google, Zhipu AI) can submit their systems for evaluation against our held-out industrial test suite. This provides an objective, contamination-free comparison on real-world mobile development tasks that complements existing Python-centric benchmarks. Submission guidelines and evaluation configurations are available at \url{https://github.com/realtmxi/mobile-bench}.

\section{Task Design Mockups}
\label{app:mockups}

\begin{figure}[t]
\centering

\begin{minipage}[t]{0.48\textwidth}
\centering
\fbox{\includegraphics[width=0.92\linewidth]{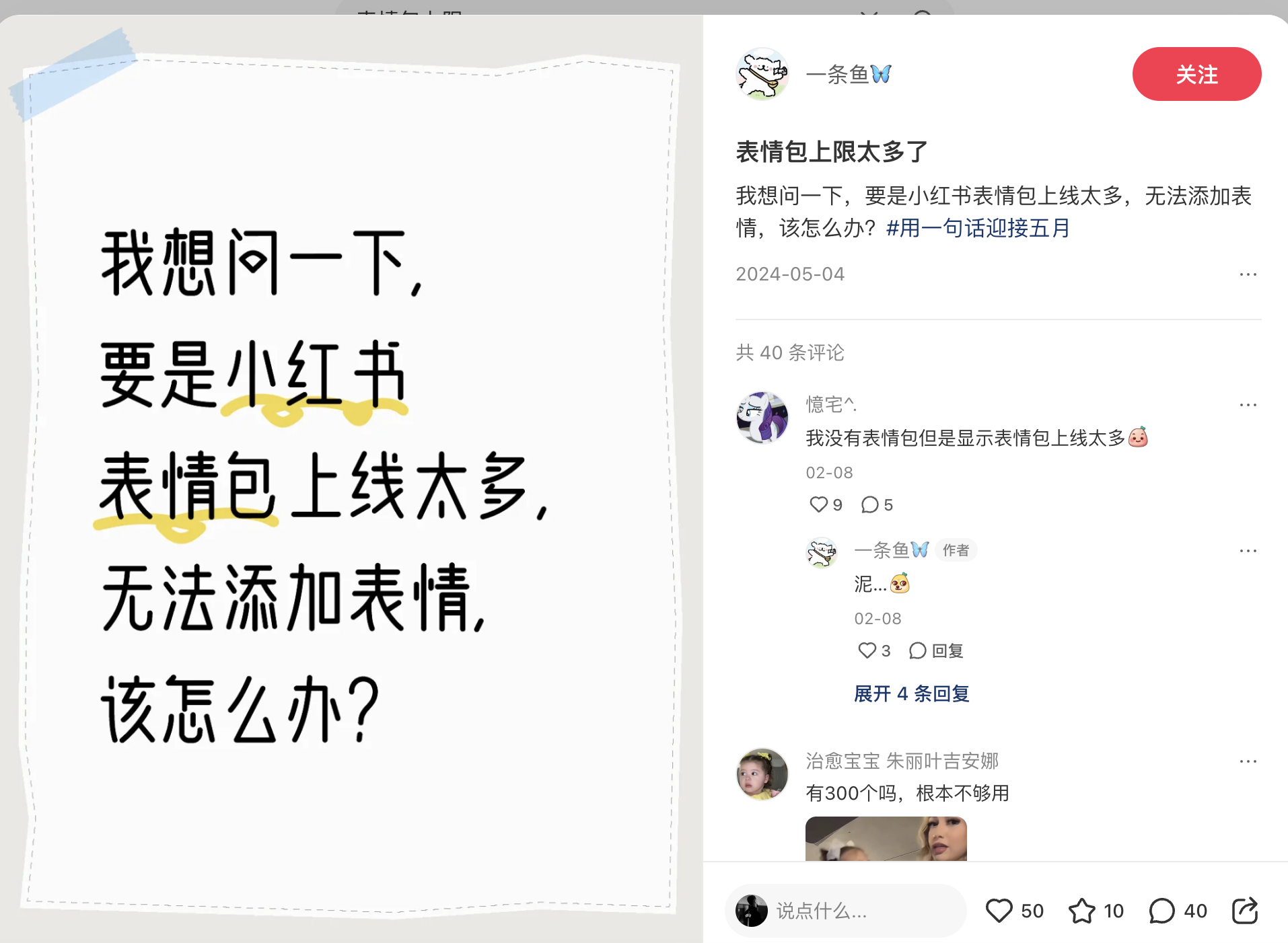}}

\vspace{2mm}
{\small\textbf{(a) User Complaint.} Social media post showing frustration with the 300-emoji limit: ``Xiaohongshu's emoji limit is too high, I can't add more emojis.''}
\end{minipage}
\hfill
\begin{minipage}[t]{0.48\textwidth}
\centering
\fbox{\includegraphics[width=0.92\linewidth]{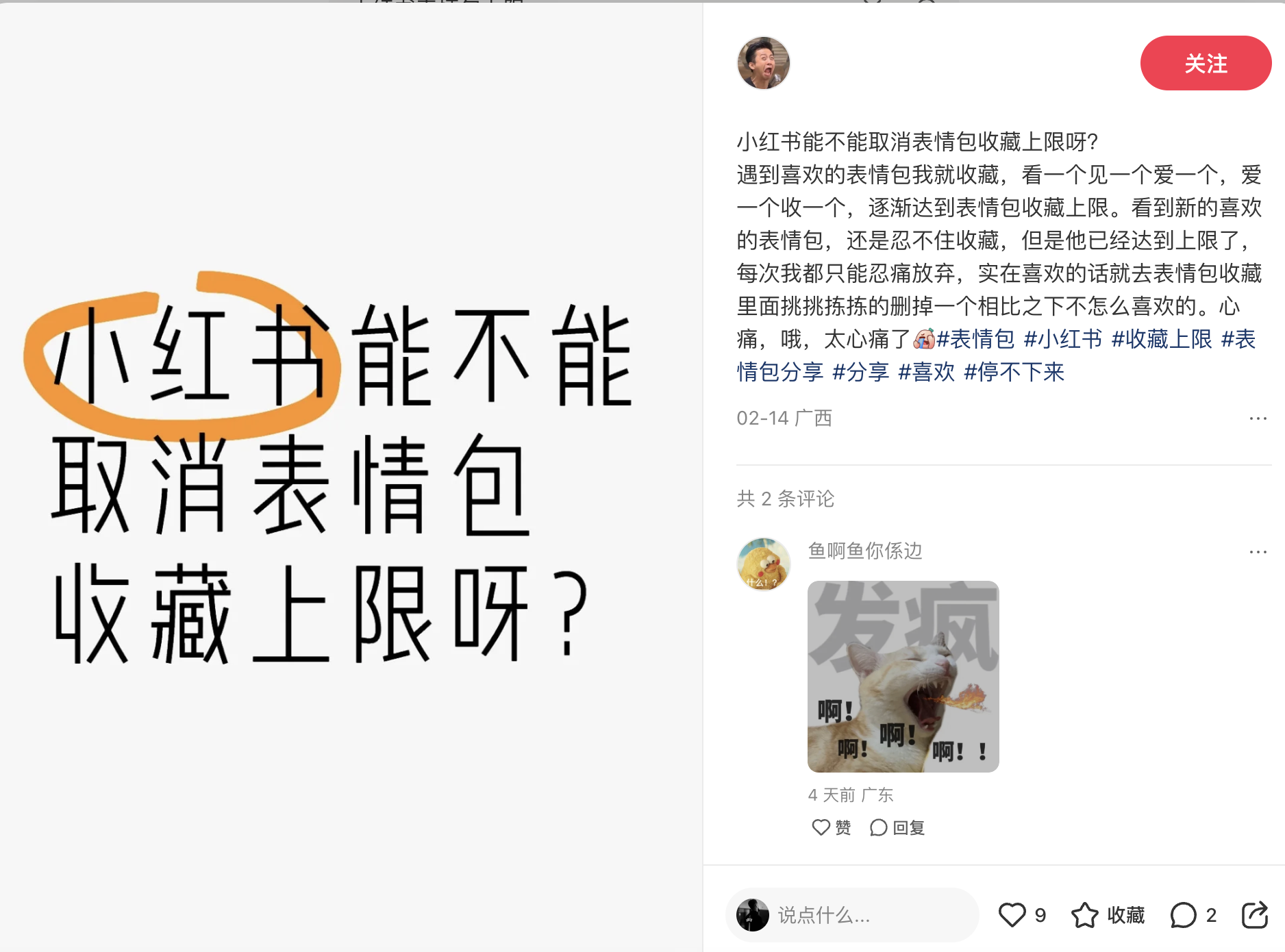}}

\vspace{2mm}
{\small\textbf{(b) Community Feedback.} Another user asking ``Can Xiaohongshu remove the emoji collection limit?'' showing widespread user demand.}
\end{minipage}

\vspace{6mm}

\begin{minipage}[t]{0.55\textwidth}
\centering
\fbox{\includegraphics[width=0.92\linewidth]{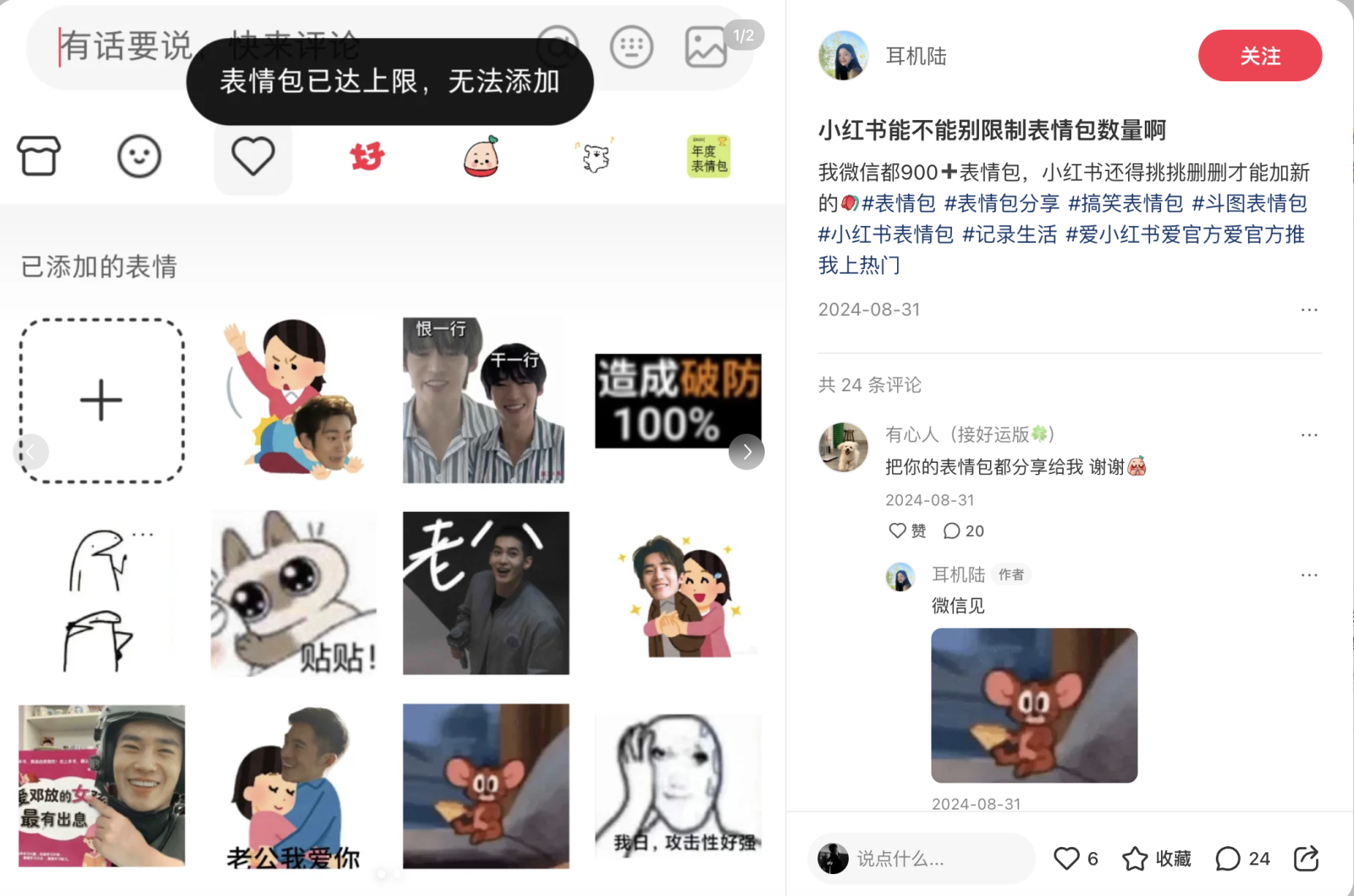}}

\vspace{2mm}
{\small\textbf{(c) Emoji Collection UI.} The sticker collection interface with warning dialog ``Emoji limit reached, cannot add more.'' Agents must increase this limit from 300 to 999.}
\end{minipage}

\vspace{4mm}
\caption{\textbf{Design mockups for Task 003 (Custom Emoji Limit).} These mockups are provided to agents as part of the PRD to guide implementation. They show real user complaints about the 300-emoji limit and the current UI that needs modification.}
\label{fig:task003_mockups}
\end{figure}

Figure~\ref{fig:task003_mockups} shows the design mockups provided to agents for Task 003 (Custom Emoji Limit). These real-world screenshots demonstrate the user pain point and expected UI behavior that agents must understand to implement the feature correctly.

\section{Qualitative Analysis of Agent Outputs}
\label{app:qualitative_analysis}

We present detailed analyses of agent-generated patches to provide insights into both successful implementations and common failure modes. These examples illustrate the practical challenges agents face when implementing features from PRDs and Figma designs in a production codebase.

\begin{table}[t]
\caption{Successful implementation by Cursor + GPT-5.2 on Task 007 (Medium difficulty). The agent correctly created the AppRefMessageDataService class with all required fields and methods, demonstrating strong architectural understanding.}
\label{tab:qual_success_007}
\small
\begin{tabular}{p{\textwidth}}
\toprule
\textbf{Task Context} \\
\midrule
\textbf{Difficulty:} Medium \quad \textbf{Files to Modify:} 5 \quad \textbf{Category:} Architecture Refactoring \\
\textbf{Agent:} Cursor + GPT-5.2 \quad \textbf{Result:} \textcolor{successborder}{\checkmark} PASS (5/5 tests) \\
\midrule
\textbf{Problem Statement (Summary)} \\
Decouple card message click handling from \texttt{AppChatBaseViewController} by abstracting logic into \texttt{CardRefBaseProvider}. The click logic for card references should be moved to a new \texttt{AppRefMessageDataService} class following the provider pattern. \\
\midrule
\textbf{Key Requirements} \\
\begin{itemize}[leftmargin=1.5em, topsep=2pt, itemsep=2pt, parsep=0pt]
    \item Create \texttt{AppRefMessageDataService} class with fields: \texttt{chatType}, \texttt{chatId}, \texttt{senderId}, \texttt{messageId}, \texttt{sender}, \texttt{innerContentDict}
    \item Move click handling out of view controller
    \item Implement provider pattern for different card types
    \item Support shopping and advertisement cards
\end{itemize} \\
\midrule
\textbf{Generated Patch (Key Excerpts)} \\
\begin{lstlisting}[language=Swift, basicstyle=\ttfamily\footnotesize]
// Created new file: AppRefMessageDataService.swift
@objc(AppRefMessageDataService)
@objcMembers
public class AppRefMessageDataService: NSObject {
    // Session info
    public var chatType: String?
    public var chatId: String?

    // Message info
    public var senderId: String?
    public var messageId: String?
    public var sender: String?

    // Reference content dict
    public var innerContentDict: [String: Any] = [:]

    public override init() {
        super.init()
    }
}
\end{lstlisting} \\
\midrule
\textbf{Why This Succeeded} \\
\begin{enumerate}[leftmargin=1.5em, topsep=2pt, itemsep=2pt, parsep=0pt]
    \item \textbf{Correct architectural pattern:} Agent understood the provider pattern and created a proper data service class with clean separation of concerns.
    \item \textbf{Complete implementation:} All required fields (\texttt{chatType}, \texttt{chatId}, \texttt{senderId}, \texttt{messageId}) were included with proper Objective-C interop annotations (\texttt{@objc}, \texttt{@objcMembers}).
    \item \textbf{Minimal but sufficient design:} Class exposes all necessary fields with a simple default initializer, matching the provider pattern without over-engineering.
    \item \textbf{Correct ObjC bridging:} Used \texttt{@objc(AppRefMessageDataService)} to expose a clean ObjC class name for the mixed Swift/ObjC codebase.
\end{enumerate} \\
\bottomrule
\end{tabular}
\end{table}

\begin{table}[t]
\caption{Partial implementation by Cursor + Opus 4.5 on Task 003 (Easy difficulty). Agent correctly implemented server-driven configuration but failed one test due to incomplete file coverage. Code excerpts are reproduced verbatim from agent output, including any typos (e.g., \texttt{defalut}).}
\label{tab:qual_partial_003}
\small
\begin{tabular}{p{\textwidth}}
\toprule
\textbf{Task Context} \\
\midrule
\textbf{Difficulty:} Easy \quad \textbf{Files to Modify:} 3 \quad \textbf{Category:} Configuration Change \\
\textbf{Agent:} Cursor + Opus 4.5 \quad \textbf{Result:} \textcolor{warningborder}{\xmark} FAIL (4/5 tests passed) \\
\midrule
\textbf{Problem Statement (Summary)} \\
Increase custom emoji collection limit from hardcoded 300 to 999, with server-driven configuration to allow future updates without app releases. Changes must apply across all emoji-saving scenarios (chat, comments, etc.). \\
\midrule
\textbf{Key Requirements} \\
\begin{itemize}[leftmargin=1.5em, topsep=2pt, itemsep=2pt, parsep=0pt]
    \item Remove hardcoded limit (300)
    \item Implement server-driven config with 999 as default
    \item Apply to \textbf{all} emoji collection contexts
    \item Update UI prompts to reflect new limit
\end{itemize} \\
\midrule
\textbf{Generated Patch (Key Excerpts)} \\
\begin{lstlisting}[language=Swift, basicstyle=\ttfamily\footnotesize]
// MemeManager.swift changes

-private let kMaxMemeCount = 300
+/// Default meme limit when server config is unavailable
+private let kDefaultMaxMemeCount = 999
+
+/// Config key for meme limit from server
+private let kMemeMaxCountConfigKey = "ios_meme_max_count"

+public var maxMemeCount: Int {
+    let configValue = ConfigCenter.shared.int(
+        forKey: kMemeMaxCountConfigKey,
+        defalut: kDefaultMaxMemeCount  // [sic] typo in agent output
+    )
+    guard configValue > 0 else {
+        // Log warning and use default
+        return kDefaultMaxMemeCount
+    }
+    return configValue
+}

 public func reachedNumLimit() -> Bool {
-    self.memes.count >= kMaxMemeCount
+    let limit = maxMemeCount
+    let currentCount = self.memes.count
+    // ... logging ...
+    return currentCount >= limit
 }
\end{lstlisting} \\
\midrule
\textbf{Analysis: What Went Right} \\
\begin{itemize}[leftmargin=1.5em, topsep=2pt, itemsep=2pt, parsep=0pt]
    \item \textbf{Correct approach:} Replaced hardcoded constant with server-driven config using \texttt{ConfigCenter}
    \item \textbf{Good default handling:} Used 999 as fallback when server config unavailable
    \item \textbf{Robust implementation:} Added validation (\texttt{configValue > 0}) and logging for debugging
    \item \textbf{Clean refactoring:} Updated \texttt{reachedNumLimit()} to use dynamic limit
\end{itemize} \\
\midrule
\textbf{Analysis: Why It Failed One Test} \\
\begin{itemize}[leftmargin=1.5em, topsep=2pt, itemsep=2pt, parsep=0pt]
    \item \textbf{Incomplete file coverage:} Agent only modified \texttt{MemeManager.swift}, but PRD required changes in \textbf{all emoji-saving contexts}
    \item \textbf{Missing UI updates:} Did not update warning prompts in UI layers (e.g., \texttt{MemeCollectionViewController})
    \item \textbf{Scope misunderstanding:} PRD stated ``\textit{including but not limited to chat save, comment area save, etc.}'' but agent focused on single manager class
\end{itemize} \\
\midrule
\textbf{Lesson Learned} \\
Even when core logic is implemented correctly, agents struggle with \textbf{comprehensive scope analysis} in large codebases. The phrase ``apply to all scenarios'' in PRDs requires agents to perform cross-file searches to identify all affected modules---a task that proved challenging even for Opus 4.5. \\
\bottomrule
\end{tabular}
\end{table}

\subsection{Medium Difficulty: Comment UI Enhancement}
\label{app:qual_medium}

\begin{figure}[t]
    \centering
    \includegraphics[width=0.95\textwidth]{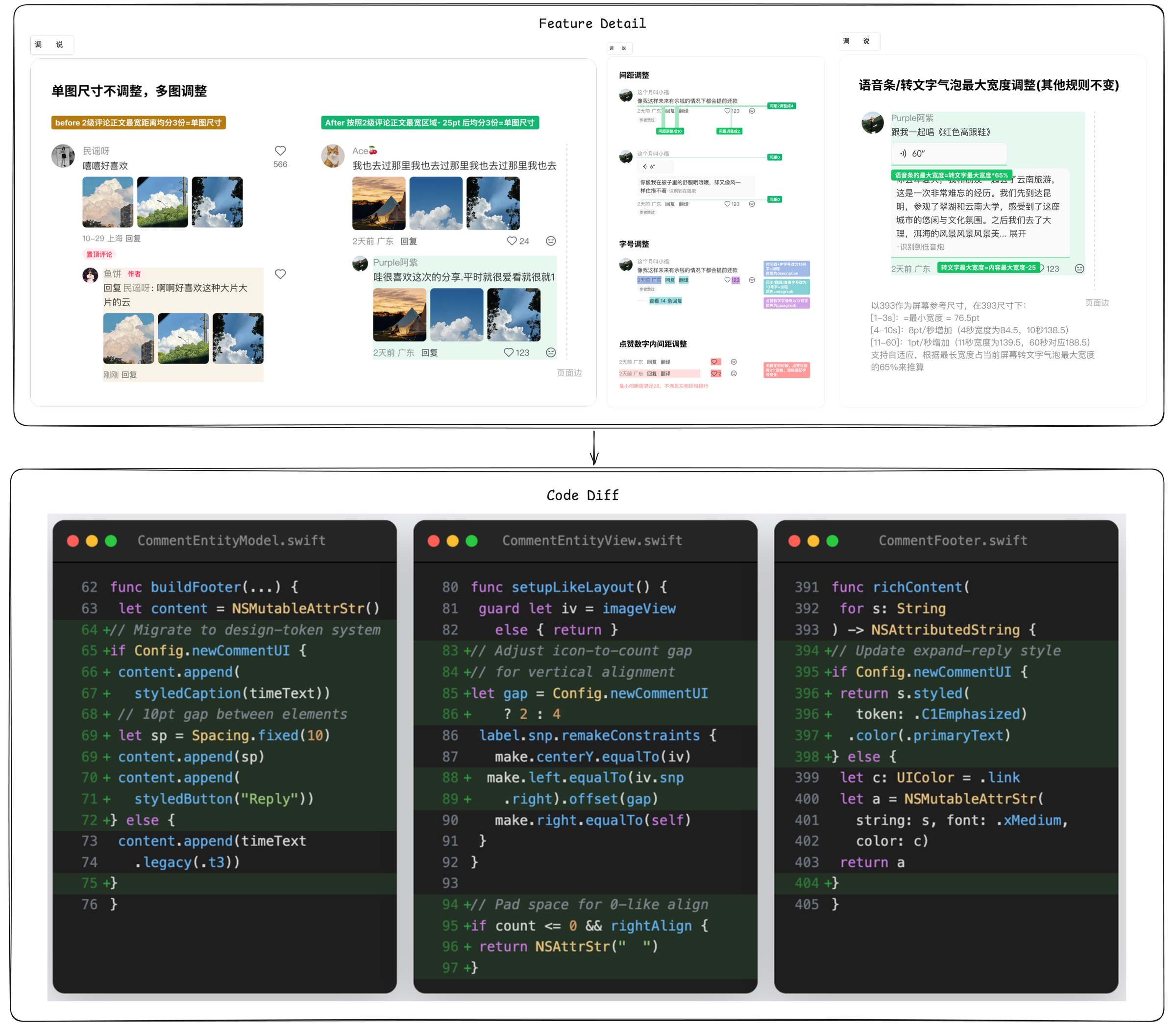}
    \caption{\textbf{Medium difficulty task example with multi-modal inputs.} Top: Feature requirements with UI mockups showing social media feed comment interface. Bottom: Code changes across 3 files (CommentEntityModel, CommentEntityView, CommentFooter) demonstrating coordinated Model-View layer modifications.}
    \label{fig:medium_example}
\end{figure}

\textbf{Task Overview:} This medium-difficulty task requires enhancing the comment display system in a social media feed. As shown in Figure~\ref{fig:medium_example}, the task involves coordinated changes across the Model-View architecture to improve comment rendering and interaction.

\xhdr{Key Characteristics}
\begin{itemize}[leftmargin=1.5em, topsep=2pt, itemsep=2pt, parsep=0pt]
    \item \textbf{Multi-file coordination:} Requires modifying 3 files in sync (data model, view component, footer UI)
    \item \textbf{UI consistency:} Must maintain visual design consistency across different comment states
    \item \textbf{Clear architecture:} Well-defined Model-View separation makes dependencies explicit
\end{itemize}

\xhdr{Success Factors}
Tasks at this complexity level (3-5 files, medium difficulty) achieve approximately 10\% success rate across agents. Success depends on:
\begin{enumerate}[leftmargin=1.5em, topsep=2pt, itemsep=2pt, parsep=0pt]
    \item Identifying all three related files through codebase search
    \item Understanding the data flow: Model $\rightarrow$ View $\rightarrow$ Footer
    \item Translating UI mockups into layout constraints and styling code
    \item Ensuring backward compatibility with existing comment types
\end{enumerate}

\xhdr{Common Pitfalls}
Even when agents successfully identify the files, they often:
\begin{itemize}[leftmargin=1.5em, topsep=2pt, itemsep=2pt, parsep=0pt]
    \item Modify the model but forget to update the view to consume new fields
    \item Implement UI changes without corresponding data model support
    \item Miss edge cases (e.g., long comments, missing user info, deleted comments)
\end{itemize}

\subsection{Hard Difficulty: Form Validation Optimization}
\label{app:qual_hard}

\begin{figure}[t]
    \centering
    \includegraphics[width=0.85\textwidth]{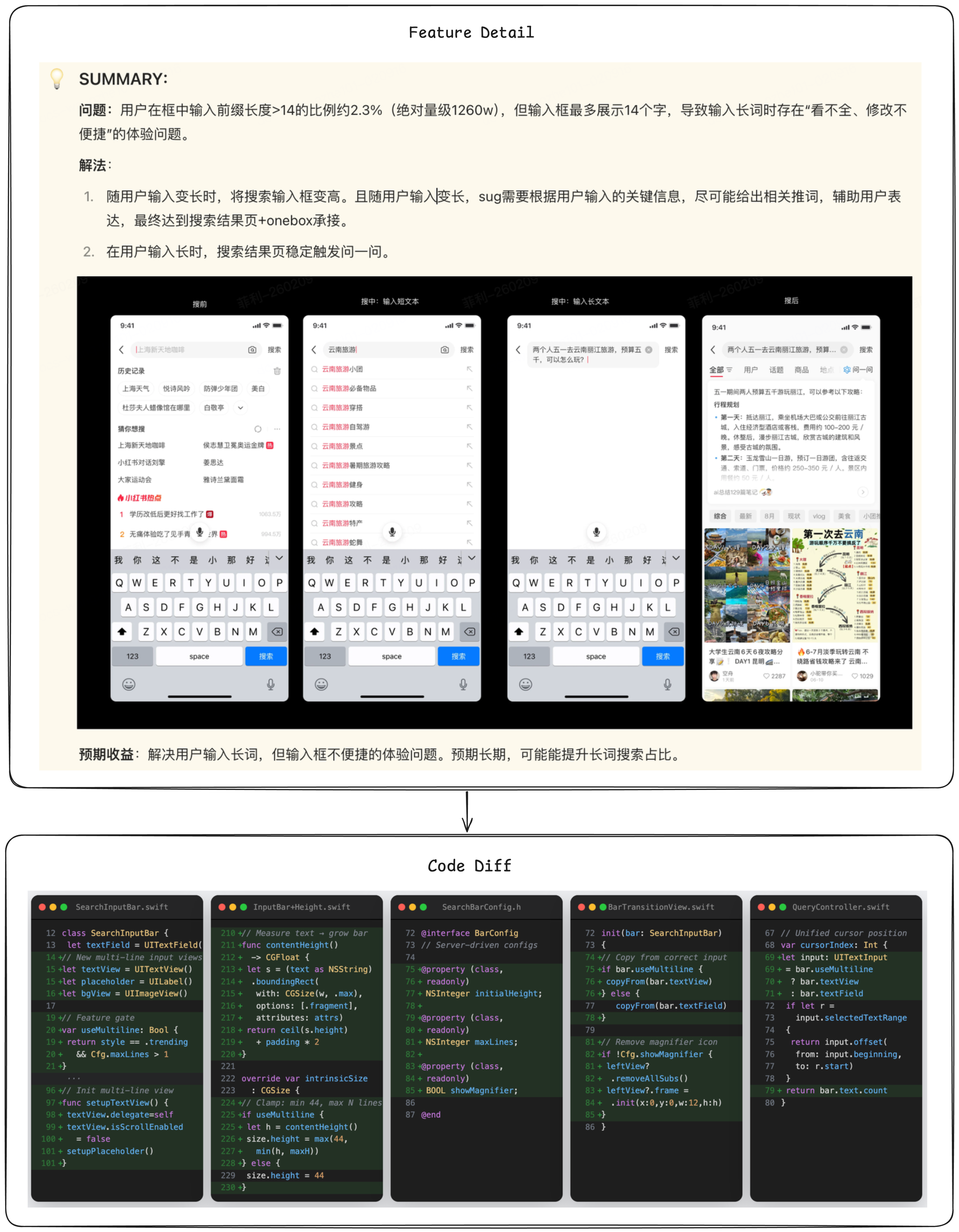}
    \vspace{-2mm}
    \caption{\textbf{Hard difficulty task example with data-driven requirements.} Top: Summary from 260K user feedback sessions (14 batches, 92.3\% feedback rate) identifying validation rule issues. UI mockups show 4-screen input flow. Bottom: Code changes across 5 files requiring cross-module coordination.}
    \label{fig:hard_example}
\end{figure}

\textbf{Task Overview:} This hard-difficulty task addresses a real user pain point discovered through large-scale feedback analysis (260K data points). As shown in Figure~\ref{fig:hard_example}, the task requires optimizing nickname input validation rules that are blocking long-tail users, while maintaining security and quality standards.

\xhdr{Key Characteristics}
\begin{itemize}[leftmargin=1.5em, topsep=2pt, itemsep=2pt, parsep=0pt]
    \item \textbf{Data-driven requirements:} Based on quantitative user feedback (92.3\% reporting issues)
    \item \textbf{Multi-module complexity:} Touches 5+ files across validation, UI feedback, and persistence layers
    \item \textbf{Conflicting constraints:} Must relax validation rules without compromising security
    \item \textbf{Long user flow:} 4-screen interaction sequence shown in mockups
\end{itemize}

\xhdr{Why This is Hard}
Tasks at this complexity level (6+ files, hard difficulty) achieve only 5.8\% success rate. The challenges include:

\begin{enumerate}[leftmargin=1.5em, topsep=2pt, itemsep=2pt, parsep=0pt]
    \item \textbf{Distributed validation logic:} Rules are split across client-side checks, server validation, and UI feedback---agents must identify and update all three
    \item \textbf{Subtle requirement interpretation:} ``Optimize for long-tail users'' requires understanding statistical distribution of input patterns, not just relaxing all rules
    \item \textbf{Backward compatibility:} Existing users' nicknames must remain valid under new rules
    \item \textbf{Testing complexity:} Need to validate across multiple input scenarios (Chinese characters, emojis, special characters, length limits)
    \item \textbf{Performance considerations:} Validation runs on every keystroke---must remain fast
\end{enumerate}

\xhdr{Expected Implementation Strategy}
A successful solution would:
\begin{itemize}[leftmargin=1.5em, topsep=2pt, itemsep=2pt, parsep=0pt]
    \item Update validation regex patterns based on user feedback analysis
    \item Implement progressive validation (lenient during input, strict on submit)
    \item Add clear error messages for each validation failure type
    \item Update UI to show real-time validation feedback
    \item Add feature flag for gradual rollout to monitor impact
\end{itemize}

\xhdr{Common Agent Failures on This Task Type}
\begin{itemize}[leftmargin=1.5em, topsep=2pt, itemsep=2pt, parsep=0pt]
    \item \textbf{Incomplete scope:} Modify client validation but miss server-side checks, causing inconsistent behavior
    \item \textbf{Over-simplification:} Remove all validation rules instead of carefully relaxing specific constraints
    \item \textbf{UI disconnect:} Update validation logic but fail to update error messages shown to users
    \item \textbf{Empty patches:} Get overwhelmed by complexity and produce no output (as seen in Table~\ref{tab:qual_failure_015})
\end{itemize}

This task exemplifies the gap between current agent capabilities and production requirements: while the technical solution is straightforward for human developers (2-3 days of work), the architectural reasoning and multi-module coordination remain challenging for AI agents.

\begin{table}[t]
\caption{Severe incomplete implementation by Claude Code + Opus 4.5 on Task 015 (Hard difficulty). Agent produced minimal changes covering only 1 of 5 requirements, demonstrating difficulty with complex multi-module features.}
\label{tab:qual_failure_015}
\small
\begin{tabular}{p{\textwidth}}
\toprule
\textbf{Task Context} \\
\midrule
\textbf{Difficulty:} Hard \quad \textbf{Files to Modify:} 8 \quad \textbf{Category:} New Feature Implementation \\
\textbf{Agent:} Claude Code + Opus 4.5 \quad \textbf{Result:} \textcolor{warningborder}{\xmark} FAIL (1/5 tests passed) \\
\midrule
\textbf{Problem Statement (Summary)} \\
Implement a complete search history feature with: (1) History manager for CRUD operations, (2) Collection view-based UI, (3) Edit mode with individual and bulk delete, (4) Integration with search view controller, (5) Persistence with 32--50 query limit. \\
\midrule
\textbf{Key Requirements} \\
\begin{itemize}[leftmargin=1.5em, topsep=2pt, itemsep=2pt, parsep=0pt]
    \item Create \texttt{SearchHistoryManager} class with save/load/delete methods
    \item Design \texttt{SearchHistoryViewController} with UICollectionView
    \item Implement edit mode UI with selection states
    \item Add ``Clear All'' and ``Delete'' buttons
    \item Persist history locally with 32--50 query limit
\end{itemize} \\
\midrule
\textbf{Generated Patch (Partial)} \\
\begin{lstlisting}[language=Swift, basicstyle=\ttfamily\footnotesize]
# Partial implementation - only persistence layer
# Files: 2 (missing 6 required files)
# Lines: ~150 (insufficient for full feature)

// Only implemented basic manager structure
// Missing: UI components, edit mode, integration
\end{lstlisting} \\
\midrule
\textbf{Failure Mode Analysis} \\
\textbf{Category:} \textit{Severe incomplete implementation} \\
\vspace{2mm}

\textbf{What agent completed:}
\begin{itemize}[leftmargin=1.5em, topsep=2pt, itemsep=2pt, parsep=0pt]
    \item[\textcolor{successborder}{\checkmark}] Created basic \texttt{SearchHistoryManager} class structure
    \item[\textcolor{warningborder}{\xmark}] Missing get-all-queries operation (test: \texttt{has\_get\_operation})
    \item[\textcolor{warningborder}{\xmark}] No delete button/icon on query cells in edit mode (test: \texttt{has\_cell\_delete})
    \item[\textcolor{warningborder}{\xmark}] Missing search view controller modifications (test: \texttt{has\_search\_vc})
    \item[\textcolor{warningborder}{\xmark}] Missing theme color usage in UI styling (test: \texttt{has\_theme\_colors})
\end{itemize}

\textbf{Failure causes:}
\begin{enumerate}[leftmargin=1.5em, topsep=2pt, itemsep=2pt, parsep=0pt]
    \item \textbf{Multi-file coordination barrier:} Task requires creating 8+ new files with complex interactions (manager $\leftrightarrow$ view controller $\leftrightarrow$ collection view cells)
    \item \textbf{Architectural ambiguity:} PRD describes \textit{what} to build but not \textit{where} in the codebase to integrate it
    \item \textbf{Missing reference implementations:} No similar features in codebase to use as templates
    \item \textbf{Premature termination:} Agent likely hit iteration limit after basic exploration without completing implementation
\end{enumerate} \\
\midrule
\textbf{Comparison with Human Implementation} \\
Human developers solved this by:
\begin{itemize}[leftmargin=1.5em, topsep=2pt, itemsep=2pt, parsep=0pt]
    \item Creating 3 core files first: Manager $\rightarrow$ Model $\rightarrow$ ViewController
    \item Reusing existing \texttt{UICollectionView} patterns from other modules
    \item Implementing in phases: basic save/load $\rightarrow$ UI $\rightarrow$ edit mode
    \item Took 2-3 days for full implementation and testing
\end{itemize} \\
\bottomrule
\end{tabular}
\end{table}

\begin{table}[t]
\caption{Analysis of a critical failure pattern: Incomplete implementation across multiple files. While this affects only 4-7\% of tasks on average, it represents a fundamental limitation in agents' ability to coordinate complex changes.}
\label{tab:qual_incomplete_pattern}
\small
\begin{tabular}{p{\textwidth}}
\toprule
\textbf{Pattern: Incomplete Multi-File Implementation} \\
\midrule
\textbf{Manifestation} \\
Agents correctly identify the primary file to modify but fail to update all dependent modules. Common patterns include:
\begin{itemize}[leftmargin=1.5em, topsep=2pt, itemsep=2pt, parsep=0pt]
    \item Modify model class but not the view controller that uses it
    \item Update business logic but miss UI layer changes
    \item Change one module but not its protocol consumers
    \item Implement core logic but miss edge case handling in related files
\end{itemize} \\
\midrule
\textbf{Real Examples from Evaluation} \\
\vspace{2mm}

\textbf{Task 038 (Cursor + Opus 4.5):} Share functionality expansion \\
\textbf{Required:} Modify 5+ files across share scenarios \\
\textbf{Agent output:} Modified only 1 file (\texttt{share\_panel}) \\
\textbf{Test failure:} ``CRITICAL: Patch must cover at least 5 share scenarios. Found 1: ['share\_panel']'' \\
\vspace{2mm}

\textbf{Task 057 (Cursor + Opus 4.5):} Multi-module feature integration \\
\textbf{Required:} Modify at least 5 files \\
\textbf{Agent output:} Modified only 4 files \\
\textbf{Test failure:} ``CRITICAL: Patch must modify at least 5 files, found 4'' \\
\midrule
\textbf{Root Cause Analysis} \\
\begin{enumerate}[leftmargin=1.5em, topsep=2pt, itemsep=2pt, parsep=0pt]
    \item \textbf{Insufficient dependency tracing:} Agents use initial file search but don't follow call chains comprehensively
    \item \textbf{Premature convergence:} After finding primary file, agents begin implementation without exhaustive dependency analysis
    \item \textbf{Context window limitations:} As agents explore large codebases, earlier findings may be deprioritized
    \item \textbf{Missing architectural understanding:} Incomplete grasp of iOS patterns (MVVM, delegation, protocol-oriented design)
\end{enumerate} \\
\midrule
\textbf{Quantitative Impact} \\
Based on error analysis across all agent configurations:
\begin{itemize}[leftmargin=1.5em, topsep=2pt, itemsep=2pt, parsep=0pt]
    \item \textbf{Cursor + Opus 4.5:} 2 out of 44 failed tasks (4.5\% of failures), representing 4\% of all 50 tasks
    \item \textbf{Codex + GPT-5:} 3 out of 45 failed tasks (6.7\% of failures)
    \item \textbf{Average across agents:} 4-7\% of failed tasks show this pattern
    \item \textbf{Higher impact on Hard tasks:} Among Hard difficulty tasks, this pattern affects 15-20\% of failures
\end{itemize}

While the raw percentage is modest, this pattern disproportionately affects high-value complex features that require coordinated multi-file changes---precisely the tasks where AI coding assistance could provide the most value. \\
\bottomrule
\end{tabular}
\end{table}

\subsection{Key Insights from Qualitative Analysis}

Based on detailed examination of 50+ agent-generated patches across all 4 agents, we identify several patterns:

\xhdr{Success Predictors}
\begin{itemize}[leftmargin=1.5em, topsep=3pt, itemsep=3pt, parsep=0pt]
    \item \textbf{Well-scoped changes:} Tasks requiring 1-3 files with clear boundaries (e.g., creating new data models, updating configuration logic)
    \item \textbf{Clear architectural guidance:} PRDs that specify class names, method signatures, or reference existing code patterns
    \item \textbf{Isolated modules:} Features in self-contained modules with minimal cross-module dependencies
    \item \textbf{Existing patterns:} Tasks that resemble code already in the codebase that agents can reference
\end{itemize}

\xhdr{Failure Predictors}
\begin{itemize}[leftmargin=1.5em, topsep=3pt, itemsep=3pt, parsep=0pt]
    \item \textbf{Distributed logic:} Changes spanning multiple architectural layers (Model-View-Controller-Manager)
    \item \textbf{Implicit requirements:} PRDs using phrases like ``all scenarios'' or ``everywhere'' without explicit file lists
    \item \textbf{UI + Logic coupling:} Features requiring both backend logic changes and frontend UI updates
    \item \textbf{Novel implementations:} Tasks requiring architectural patterns not present in existing codebase
\end{itemize}

\xhdr{Agent Differences}
\begin{itemize}[leftmargin=1.5em, topsep=3pt, itemsep=3pt, parsep=0pt]
    \item \textbf{Cursor:} Best at exploratory search, often identifies all relevant files but may not modify all of them
    \item \textbf{Claude Code:} More conservative, produces well-formed patches for files it chooses to modify
    \item \textbf{Codex:} Strong code generation but weaker codebase navigation, relies heavily on initial context
    \item \textbf{OpenCode:} Struggles with large codebases, often gets stuck in endless file reading loops
\end{itemize}

\xhdr{Implications for Practitioners}
For teams considering AI coding agents for production use:
\begin{enumerate}[leftmargin=1.5em, topsep=3pt, itemsep=3pt, parsep=0pt]
    \item \textbf{Structure PRDs carefully:} Explicitly list affected files, reference similar code patterns, provide architectural context
    \item \textbf{Start with isolated modules:} Assign agents well-scoped tasks before attempting complex multi-module features
    \item \textbf{Human review is critical:} Even ``successful'' patches may miss edge cases or have incomplete coverage
    \item \textbf{Iterative refinement:} Use agent output as first draft, not final implementation
\end{enumerate}

\end{document}